\documentclass{aa}

\usepackage{multirow}
\usepackage{graphicx}
\usepackage[varg]{txfonts}
\usepackage{natbib}
\usepackage[pdftex,colorlinks=true,linkcolor=blue,citecolor=blue,urlcolor=blue]{hyperref}

\def\A{\mathcal{A}}
\def\B{\mathcal{B}}
\def\K{\mathcal{K}}
\def\T{\mathcal{T}}
\def\Z{\mathcal{Z}}
\def\b{{b}}
\def\M{M}
\def\m{m}
\def\R{R}
\def\f{f}
\def\dgam{2 (\theta-\f)}
\def\omg{\omega}
\def\lbf{\upsilon}
\def\gam{\gamma}
\def\im{i}
\def\expo#1{\mathbf{e}^{#1}}

\def\Xh{\mathcal{X}}

\def\C22i{C_{22}}

\def\kf{k_\mathrm{f}}

\def\taua{\tau_e}
\def\taub{\tau}
\def\vect#1{{\boldsymbol{#1}}}
\def\be{\begin{equation}}
\def\ee{\end{equation}}

\def\bibpath{/Users/acorreia/Dropbox/Saved/Articles/}
\def\figpath{}
\def \llabel#1{\label{#1}}
\begin{document}

\title{Spin-orbit coupling for close-in planets}
\titlerunning{Spin-orbit coupling for close-in planets}

\author{
Alexandre C. M. Correia\inst{1,2}
\and 
Jean-Baptiste Delisle\inst{2,3}
}

\authorrunning{A.C.M. Correia \& J.-B. Delisle}


\institute{
CFisUC, Department of Physics, University of Coimbra, 3004-516 Coimbra, Portugal
\and 
ASD, IMCCE, Observatoire de Paris, PSL Universit\'e, 77 Av. Denfert-Rochereau, 75014 Paris, France
\and
Observatoire de l'Universit\'e de Gen\`eve, 51 chemin des Maillettes, 1290 Sauverny, Switzerland
 }

\date{\today}

\abstract{
We study the spin evolution of close-in planets in multi-body systems and present a very general formulation of the spin-orbit problem. This includes a simple way to probe the spin dynamics from the orbital perturbations, a new method for computing forced librations and tidal deformation, and general expressions for the tidal torque and capture probabilities in resonance. We show that planet-planet perturbations can drive the spin of Earth-size planets into asynchronous or chaotic states, even for nearly circular orbits. We apply our results to Mercury and to the KOI-1599 system of two super-Earths in a 3/2 mean motion resonance. 
}

\keywords{celestial mechanics -- planets and satellites: general -- chaos}

\maketitle


\section{Introduction}
\llabel{intro}

The inner planets of the solar system, like the majority of the main satellites, currently present a rotation state that is different from what is believed to have been the initial state \citep[e.g.,][]{Goldreich_Soter_1966}.
Just after their formation, they are supposed to have rotated much faster \citep[e.g.,][]{Kokubo_Ida_2007}, but subsequently slowed down owing to tidal interactions with the central more massive body \citep[e.g.,][]{MacDonald_1964}.

The last stage for tidal evolution is synchronous rotation with the orbital mean motion, since it corresponds to the minimum of energy \citep{Hut_1980, Adams_Bloch_2015}.
This equilibrium, also known as the 1/1 spin-orbit resonance, was first observed for the Moon.
However, other possibilities exist, such as the 3/2 spin-orbit resonance of Mercury \citep{Pettengill_Dyce_1965}, which is possible because the eccentricity is different from zero \citep{Colombo_1965,Goldreich_Peale_1966,Correia_Laskar_2004, Noyelles_etal_2014}. 
Another interesting example is the chaotic rotation of Hyperion \citep[][]{Wisdom_etal_1984, Harbison_etal_2011}, which also results from nonzero eccentricity.

When the eccentricity is nearly zero, asynchronous spin-orbit resonances are still possible, provided that a companion is present in the system.
The rotation of Venus is also tidally evolved and not synchronous \citep{Goldstein_1964, Carpenter_1964};
it is close to a -5/1 spin-orbit resonance with the Earth, which suggests that the rotation can be trapped at this ratio\footnote{The expected libration width in resonance does not seem sufficient to cover this scenario. The current explanation is that Venus rotation is an equilibrium between gravitational and thermal atmospheric tides \citep{Gold_Soter_1969, Dobrovolskis_1980, Correia_Laskar_2001}.} \citep{Goldreich_Peale_1966, Goldreich_Peale_1967}.
Indeed, one possibility for asynchronous rotation is a spin-orbit resonance with a conjunction frequency.
The minor satellites of the Pluto-Charon system may also have ended up trapped in a resonance of this kind \citep{Correia_etal_2015} or even present chaotic rotation \citep{Showalter_Hamilton_2015}. 
Another possibility, which is very important for planets involved in mean motion resonances, is resonant rotation with the orbital libration frequency \citep{Correia_Robutel_2013, Leleu_etal_2016, Delisle_etal_2017}.

Despite the proximity of solar system bodies, the determination of their rotational periods has only been achieved in the second half of the $20^\mathrm{th}$ century. 
Mercury and Venus using radar ranging \citep{Goldstein_1964, Carpenter_1964, Pettengill_Dyce_1965}, Hyperion after the Voyager~2 flyby \citep{Smith_etal_1982}, and the minor satellites of the Pluto with images from the Hubble Space Telescope \citep{Showalter_Hamilton_2015}.
We thus do not expect that it will be easy to observe the rotation of recently discovered exoplanets.
Nevertheless, many of these planets are close to their host star, and we can assume that their spins have already undergone enough dissipation and evolved into a final equilibrium.
Therefore, we can try to anticipate all the possible scenarios for the rotation of these planets.
In particular, we aim to determine when asynchronous rotation is possible, which is important for habitability studies.

In this paper we generalize previous studies to any kind of orbital perturbations.
In Sect.~\ref{soc}, we write the equation for the rotation of a planet that is perturbed by other bodies, and show that the orbital forcing can be decoupled from the rotation.
In Sect.~\ref{rigid}, we determine the possible configurations for the rotation of a rigid body by analyzing the orbital forcing.
In Sect.~\ref{forced}, we consider that the planet can deform under the action of tides, and present a new method for determining the equilibrium shape and the forced oscillations in resonance.
In Sect.~\ref{capture}, using a Maxwell model, we obtain the tidal torque as a function of the orbital forcing and the corresponding capture probabilities in resonance.
In Sect.~\ref{evol}, we perform complete numerical simulations including tidal dissipation for the KOI-1599 system.
Finally, we discuss our results in Sect.~\ref{concdisc}.

\section{Orbital forcing}
\llabel{soc}

We consider a system composed by a central star of mass $\M$, a companion planet of mass $\m$, and additional companions in coplanar orbits.
The planet is considered an oblate ellipsoid with mean radius $\R$ and gravity field coefficients given by $C_{22}$ and $S_{22}$.
We furthermore assume that the spin axis of the planet is orthogonal to the orbital plane (zero obliquity).
The torque of the star on the planet changes the rotation angle, $\theta$, whose variations are given by \citep[e.g.,][]{Correia_etal_2014}:
\be
\ddot \theta = - \frac{6 G \M \m \R^2}{C r^3} \Big[ C_{22} \sin \dgam + S_{22} \cos \dgam \Big]
\ , \llabel{130104d}
\ee
where $G$ is the gravitational constant, $C$ is the moment of inertia with respect to the rotation axis, $r$ is the distance and $\f$ is the true longitude of the center of mass of the planet with respect to the star.
We neglect terms in $(\R/r)^3$ (quadrupolar approximation) and the torques of the companions on the planet's spin.

The above expression can be rewritten in a more compact and usable form with complex numbers as 
\be
  \label{eq:ddtheta}
\ddot{\theta} = \B \, \Im \left(Z \,x \, \expo{-2\im\theta}\right) \ ,
\ee
with
\be
 Z = C_{22} - \im S_{22}
\ , \llabel{190221a}
\ee
\be
 x = \left(\frac{a}{r}\right)^3 \expo{2\im \f}
\ , \llabel{190221b}
\ee
\be
\B = \frac{6 G \M \m \R^2}{C a^3} \approx 18 \, n^2 
\ , \llabel{190221c}
\ee
where 
$a$ is a scale constant with units of distance, that is usually made equal to a mean semimajor axis,
$n$ is the mean motion, and $C / \m \R^2 \approx 1 / 3$, which is the usual value for rocky planets \citep{Yoder_1995cnt}.
This way, we are able to isolate in equation (\ref{130104d}) all the important quantities that change the rotation, namely, the shape (or deformation) of the planet, $Z$, the orbital forcing, $x$, and the rotation angle, $\expo{-2\im\theta}$.
This factorization allows us to treat each contribution separately.

The orbital and spin evolutions are coupled, such that the total angular momentum is conserved.
Therefore, a change in the rotation angle (Eq.\,(\ref{130104d})) also modifies the position $r$ \citep[see Eqs.\,(15) and (16) in][]{Correia_etal_2014}.
However, the corrections in the orbital motion are on the order of $(R/r)^2 |Z| \ll 1$ (the orbital angular momentum is much larger than the rotational angular momentum), so their contribution can usually be neglected.
For planets in multi-body systems, the important contributions to the orbital forcing thus come from the mutual gravitational perturbations. 
We let
\be
X = r \, \expo{\im \f} = (r \cos \f, r \sin \f)  
\llabel{190221d}
\ee
be the position vector of the planet, whose spin we want to study.
The evolution of $X$ can be obtained by integrating a system of $N$ point-mass bodies (the star, the planet and the remaining companions).
The orbital forcing on the spin (Eq.\,(\ref{190221b})) can then be directly computed from $X$ as
\be
x = \frac{a^3 X^2}{|X|^5} 
\ . \llabel{190221e}
\ee

Assuming quasi-periodic orbital motion, we can expand the previous expression as a Fourier series
\be
  x = \sum_{\vect{k}} x_\vect{k} \, \expo{\im \vect{k}\cdot\vect{\nu} t} 
  \ , \llabel{eq:x}
\ee
where $\vect{\nu}$ is the vector of fundamental frequencies, and the amplitudes $|x_\vect{k}|$ decay with $||\vect{k}||$, such that $\sum_{\vect{k}} |x_\vect{k}|$ is convergent.
The orbital forcing on the spin, $x$, can then be characterized by the main terms in the above series.

For the shape of the planet, $Z$, we can split it in two main contributions: one rigid or permanent, $Z_0$, due to inhomogeneous mass distribution inside the planet, and another that accounts for the deformation, $Z_t$, due to external tidal forcing, such that $Z=Z_0+Z_t$.
In the solar system, we observe that $Z_0$ is a function of the planet mass.
We have $|Z_0| \sim 0.1$ for bodies with diameters smaller than 100~km \citep[e.g.,][]{Correia_2018}, but larger bodies evolve into spherical shapes ($|Z_0| \rightarrow 0$); we have $|Z_0| \sim 10^{-5}$ for Mercury \citep{Genova_etal_2013} and $|Z_0| \sim 10^{-6}$ for the Earth and Venus \citep{Yoder_1995cnt}.
On the other hand, $|Z_t| \sim \A$ (Eq.\,(\ref{eq:Ze})), that is, it depends on the mass and distance of the star.
For Mercury we have $|Z_t| \sim 10^{-7} \ll |Z_0|$, so we can assume $Z \simeq Z_0$.
However, for the Galilean satellites of Jupiter, we have $|Z_t| \sim 10^{-4} \gg |Z_0|$, so we can assume $Z \simeq Z_t$ \citep{Correia_Rodriguez_2013}.

\section{Rigid bodies}
\label{rigid}

As in most previous studies on spin-orbit coupling, let us assume for now that $Z = Z_0$, either because $|Z_t| \ll |Z_0|$ (case of Mercury) or because $Z_t$ is frozen from a time the planet was more fluid (case of the Moon).
Then, equation (\ref{eq:ddtheta}) becomes\footnote{Many authors adopt a frame where $Z_0$ is a real number ($Z_0=C_{22,0}$), which simplifies expression (\ref{130104d}).
This is correct for the rigid case, but it is no longer valid when we consider the tidal deformation of the planet.}
\be
  \label{eq:ddtheta0}
  \ddot{\theta} = \B \, \Im \left( {Z_0} \sum_{\vect{k}} x_\vect{k} \, \expo{\im (\vect{k}\cdot\vect{\nu} t - 2 \theta)} \right)
 \ .
\ee
Each term in this expression individually behaves like a pendulum, with islands of rotational libration centered at
\be
\langle \dot \theta \rangle = \nu_\vect{k} = \frac{\vect{k}\cdot\vect{\nu}}{2} \ ,
\llabel{190222a}
\ee
and libration width
\be
\beta_\vect{k} = \sqrt{2 \B |  Z_0 x_\vect{k} |} \approx 6 n \sqrt{| Z_0|}  \sqrt{|x_\vect{k} |}
\ . \llabel{190225a}
\ee
In general, the rotation can be trapped in any of these islands, provided they are far apart.
However, when the libration widths of some individual resonant islands overlap, the rotation can be chaotic \citep{Chirikov_1979, Morbidelli_2002}, i.e., the rotation exhibits random variations in short periods of time.
Overlap between two resonances with $\nu_{\vect{k}_1} < \nu_{\vect{k}_2} $ then occurs for $\nu_{\vect{k}_1} + \beta_{\vect{k}_1} \approx \nu_{\vect{k}_2} - \beta_{\vect{k}_2}$, that is, when
\be
\bar \beta = \beta_{\vect{k}_1} + \beta_{\vect{k}_2} \approx \nu_{\vect{k}_2} - \nu_{\vect{k}_1} = \Delta \nu
\ . \llabel{150618a}
\ee
This simple condition allows us to differentiate easily between the different dynamical regimes. 
Indeed, when $Z_0$ is known for a given kind of body, we can infer its dynamics simply by analyzing the orbital forcing (Eq.\,(\ref{eq:x})).
For $\bar \beta \ll \Delta \nu$ the rotation can be trapped in any isolated island with $\dot \theta = \nu_\vect{k}$, for $\bar \beta \sim \Delta \nu $ the rotation is chaotic, while for $\bar \beta \gg \Delta \nu$ the rotation can be stable again because the $\nu_\vect{k}$ are engulfed in a single libration island with only a small chaotic zone at the separatrix.


\subsection{Single planet case}
\label{spc}

In the classical case of a single planet in an elliptical orbit, the orbital forcing (Eq.\,(\ref{190221b})) is given by
\be
  x = \sum_{k} \Xh_k^{-3,2}(e) \, \expo{2\im \varpi} \expo{\im k n t} \ ,
\llabel{190417a}
\ee
where $\Xh_k^{l,m} (e)$ are the Hansen coefficients \citep{Hughes_1981} and $\varpi$ is the longitude of the pericenter.
Thus, resonant islands occur for $\nu_k = kn/2$, with amplitude $|x_k| = \Xh_k^{-3,2}(e)$.
The two largest terms correspond to $k=2$ and $k=3$, for which $\Xh_2^{-3,2}(e) \approx 1$ and $\Xh_3^{-3,2}(e) \approx 7e/2$, respectively (we neglect terms in $e^2$).
Then
\be
\bar \beta / \Delta \nu \approx 12 \sqrt{| Z_0|} \,  \Big( 1 +  \sqrt{7e/2} \Big)
\ . \llabel{190228a}
\ee
For Mercury ($|Z_0| \approx 10^{-5}$, $e=0.206$), we get $ \bar \beta / \Delta \nu \approx 0.07 \ll 1$, so we can expect stable rotation.
Indeed, the planet is observed in a stable spin-orbit resonance with $k=3$ \citep{Colombo_1965, Goldreich_Peale_1966}.
On the other hand, for Saturn's moon Hyperion ($|Z_0| \approx 0.03$, $e = 0.123$), we get $ \bar \beta / \Delta \nu \approx 3.5 \sim 1$.
According to condition (\ref{150618a}), resonance overlap occurs for Hyperion, which shows chaotic rotation \citep{Harbison_etal_2011}.

\begin{figure}
\centering
    \includegraphics[width=\columnwidth]{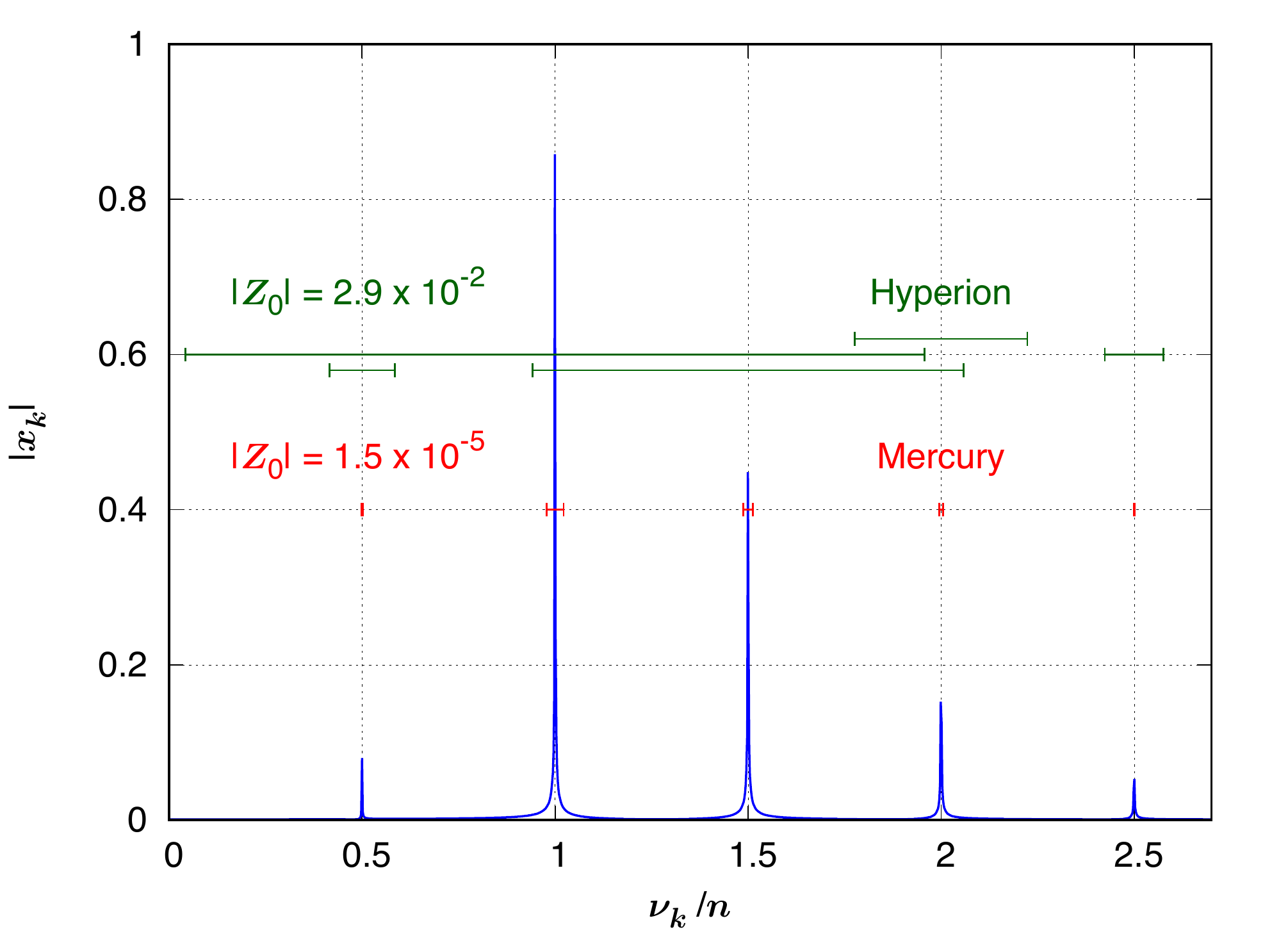}
 \caption{Fast Fourier transform of the orbital forcing $x$ (Eq.\,(\ref{190221e})) for the two-body problem with eccentricity $e=0.165$. The horizontal lines give the amplitudes $\beta_k$ for $|Z_0| = 1.5 \times 10^{-5}$, which corresponds to Mercury's value \citep{Genova_etal_2013}, and $|Z_0| = 2.9 \times 10^{-2}$, which corresponds to Hyperion's value \citep{Thomas_etal_2007}.} 
\label{FFTMH}
\end{figure}

The single planet case can be treated analytically.
However, we can also study it numerically, in order to understand more complex situations.
We can integrate the orbital motion of the two-body problem with an eccentricity $e=0.165$ (mid-value of the orbits of Mercury and Hyperion), which provide us a time series for $X$. 
We then compute a fast Fourier transform of the quantity $x$ (Eq.\,(\ref{190221e})).
The frequencies $\nu_k/n = k/2$ immediately show up as well as the corresponding amplitudes $|x_k|$.
Using expression (\ref{190225a}) we obtain the rotational libration width corresponding to the resonance islands $\dot \theta = \nu_k$ for a given value of $|Z_0|$.
In Figure~\ref{FFTMH} we horizontally draw the width of the libration islands centered at each $\nu_k$ for the $|Z_0|$ values of Mercury and Hyperion.
This plot clearly shows why Mercury can be trapped in isolated spin-orbit resonances, while Hyperion shows chaotic motion.

\subsection{Multi-planet systems}
\label{mps}

In general, planets are not alone in their orbits.
For widely spaced orbits, as in the solar system, planet-planet mutual interactions are relatively weak, and only modify the orbital parameters on secular timescales.
As a result, the single planet case from the previous section is a good approximation.
However, a large number of exoplanets has been observed in compact configurations and/or showing resonant interactions.
For these systems, the mutual perturbations are strong and need to be taken into account in the spin-orbit coupling analysis.

One possibility is to expand the disturbing function in its main perturbing terms \citep[see Chapter 6,][]{Murray_Dermott_1999}, and use it to obtain the dominating contributions to $x$ (Eq.\,(\ref{eq:x})).
Closely packed systems usually have small eccentricities, so a common approach is to truncate the series for some power in the eccentricities \citep[e.g.,][]{Goldreich_Peale_1966, Goldreich_Peale_1967, Correia_Robutel_2013, Correia_etal_2015, Leleu_etal_2016, Delisle_etal_2017}.
Although modern series manipulators allow us to perform this calculation rapidly, the number of terms to retain is large (even for a two-planet system),
and their amplitudes depend on the masses and semimajor axes. 
Moreover, for resonant systems the eccentricities can be high and the libration amplitude is also important.
Therefore, the dominating terms vary from system to system and it is difficult to predict the terms to be retained. 

A powerful alternative to analytical studies is to obtain the series expansion for $x$ numerically.
This method is exact (without approximations), fast, and valid for an unlimited number of planets in the system.
We simply need to run a simulation of the full system and then perform a frequency analysis of the orbital motion.
The only requirement is that we integrate the system for a time long enough to extract the frequencies whose amplitudes may be dominating.
However, very small frequencies (corresponding to very long periods) have no perceptible effect on the spin, since they are engulfed by the synchronous resonance (case where $\bar \beta \gg \Delta \nu$).
Thus, in general a short integration of the system ($10^4$ times the longest orbital period) is enough to characterize the spin dynamics.

As an example, we apply the numerical study to the KOI-1599 system, whose planets were detected using the transit method \citep{Steffen_Hwang_2015} and validated with transit time variations \citep{Panichi_etal_2019}.
This system is a good candidate for a rich spin dynamics because it shows some transit time variations \citep{Delisle_etal_2017}. 
The best stable fits yield a two super-Earth system in nearly circular orbits and involved in a 3/2 mean motion resonance (Table~\ref{TkoiGEAI}).
We integrated the KOI-1599 system for $1000$~years, and then performed a frequency analysis of the forcing function $x$ \citep{Laskar_1993PD}.
In Table~\ref{TkoiAF}, we show the results for the outer orbit because $\Delta \nu/n$ is larger, i.e., the individual resonant islands are more distant.
We observe that the classic perturbing terms at half-integers of the mean motion (section~\ref{spc}) are still present, but with very low amplitude, since the eccentricity is always very small ($e < 0.025$).
The three dominating terms correspond to harmonics of the mean motion, $n$, and the orbital libration frequency, $\lbf$, with similar amplitudes.

  \begin{table}
    \begin{center}
      \caption{Parameters for the KOI-1599 system used in this study.}
      \begin{tabular}{cc|cc}
        \hline
        Parameter & [unit] & inner & outer \\
        \hline
        $\m$ & [$M_\oplus$] &
        $8.3$ & $4.3$ \\
        $\R$ & [$R_\oplus$] &
        $1.9$ & 1.9 \\
        $a$ & [au] &
        $0.11230$ & $0.14727$ \\
        $e$ & &
        $0.0067$ & $0.0177$ \\
        $\lambda$ & [deg] &
        $217.4$ & $251.5$ \\
        $\varpi$ & [deg] &
        $40.6$ & $220.1$ \\
        \hline
      \end{tabular}
      \tablefoot{The mass of the star is $1.02 \, M_\odot$. This orbital solution corresponds to the stable fit GEA~I solution obtained by \citet{Panichi_etal_2019}.}
      \label{TkoiGEAI}
    \end{center}
  \end{table}

\begin{table}
\caption{Quasi-periodic decomposition of the orbital forcing function $x$ for the outer planet in the KOI-1599 system (Eq.\,(\ref{eq:x})).} 
  \begin{center}
    \begin{tabular}{rrr|rrr}
\hline
   \multicolumn{3}{c|}{$\vect{k} \cdot \vect{\nu}$} & \multicolumn{1}{c}{\multirow{2}{*}{$\nu_\vect{k} / n-1$}}  & \multicolumn{1}{c}{\multirow{2}{*}{$|x_\vect{k}|$}}  & \multicolumn{1}{c}{\multirow{2}{*}{$\varphi_\vect{k}\, (^\circ)$}} \\
\multicolumn{1}{c}{$n$} & \multicolumn{1}{c}{$\,\, \lbf$} & \multicolumn{1}{c|}{$g$}  \\ 
\hline
$2$ & $ 0$ & $ 0$ & $ 0.000000$ & $0.93142 $ & $173.21$  \\ 
$2$ & $ 1$ & $ 0$ & $ 0.003316$ & $0.25878 $ & $   79.54$  \\ 
$2$ & $-1$ & $ 0$ & $-0.003316$ & $0.24821$ & $  86.87$  \\ 
$2$ & $-2$ & $ 0$ & $-0.006632$ & $0.04461$ & $   0.54$  \\ 
$3$ & $-1$ & $-1$ & $ 0.498128$ & $0.04138$ & $ 150.57$  \\ 
$3$ & $ 0$ & $-1$ & $ 0.501445$ & $0.03730$ & $-123.10$  \\ 
$2$ & $ 2$ & $ 0$ & $ 0.006632$ & $0.02197 $ & $ -14.13$  \\ 
$3$ & $-2$ & $-1$ & $ 0.494812$ & $0.01212$ & $  64.24$  \\ 
$3$ & $ 1$ & $-1$ & $ 0.504761$ & $0.01070$ & $ 143.24$  \\ 
$1$ & $ 0$ & $ 1$ & $-0.501445$ & $0.00753$ & $ -70.49$  \\ 
$2$ & $-3$ & $ 0$ & $-0.009949$ & $0.00676$ & $ -85.79$  \\ 
  \hline
      \end{tabular}
      \tablefoot{We provide the dominating terms of $x$ by decreasing amplitude ($|x_\vect{k}| > 5 \times 10^{-3}$). We have $\nu_\vect{k} = \tfrac12 \vect{k} \cdot \vect{\nu}$ (Eq.\,(\ref{190222a})), and $\varphi_\vect{k} = \arg x_\vect{k}$. $\vect{\nu} = (n,\lbf,g)$ is the fundamental frequencies vector of the orbital motion, where $n=112.305$~rad/yr is the mean motion, $\lbf=0.744857$~rad/yr is the libration frequency, and $g=-0.324472$~rad/yr is the precession rate.} 
\label{TkoiAF}
\end{center}
\end{table}

\begin{figure}
\centering
    \includegraphics[width=\columnwidth]{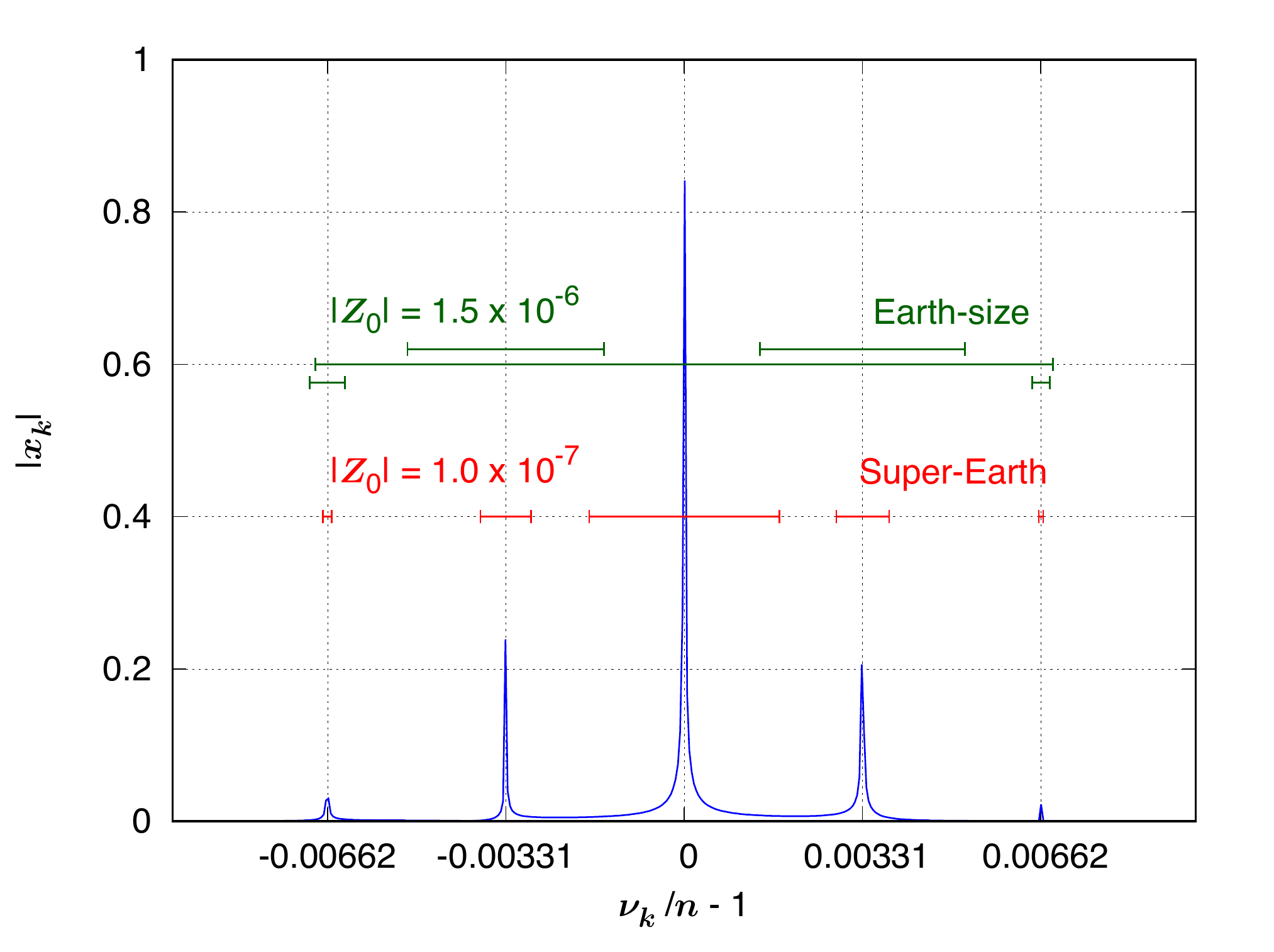}
 \caption{Fast Fourier transform of the orbital forcing $x$ (Eq.\,(\ref{190221e})) for the outer planet in the KOI-1599 system. The horizontal lines give the amplitudes $\beta_\vect{k}$ for $|Z_0| = 1.5 \times 10^{-6}$ and $|Z_0| = 1.0 \times 10^{-7}$.} 
\label{FFTKOI}
\end{figure}

We need to estimate the rotational libration width $\beta_\vect{k}$ for each frequency in Table~\ref{TkoiAF} to determine the behavior of the spin.
Since we know the orbital forcing amplitudes, the only uncertainty is related to the permanent deformation of the planet, $Z_0$ (Eq.\,(\ref{190225a})). 
We can adopt the Earth's value, $|Z_0| = 1.5 \times 10^{-6}$ because KOI-1599 planets are likely rocky, as their densities are similar to that of the Earth.
However, their surface gravity is a bit stronger, so maybe the inhomogeneities in the mass distribution are smaller, such that $|Z_0| \sim 10^{-7}$.
These two values can be seen as maximal and minimal estimations for $|Z_0|$, respectively. 
In Figure~\ref{FFTKOI} we horizontally draw the libration width of the islands centered at each $\nu_\vect{k}$ for both $|Z_0|$ values over the Fourier spectra of the orbital forcing $x$.
We observe that, for $|Z_0| \sim 10^{-7}$, each island is isolated, so the rotation can be trapped in any of them, which includes the synchronous resonance, along with the super- and sub-synchronous resonances.
On the other hand, for $|Z_0| \sim 10^{-6}$, the libration widths overlap, so we can expect chaotic rotation.

\begin{figure}
\centering
    \includegraphics[width=\columnwidth]{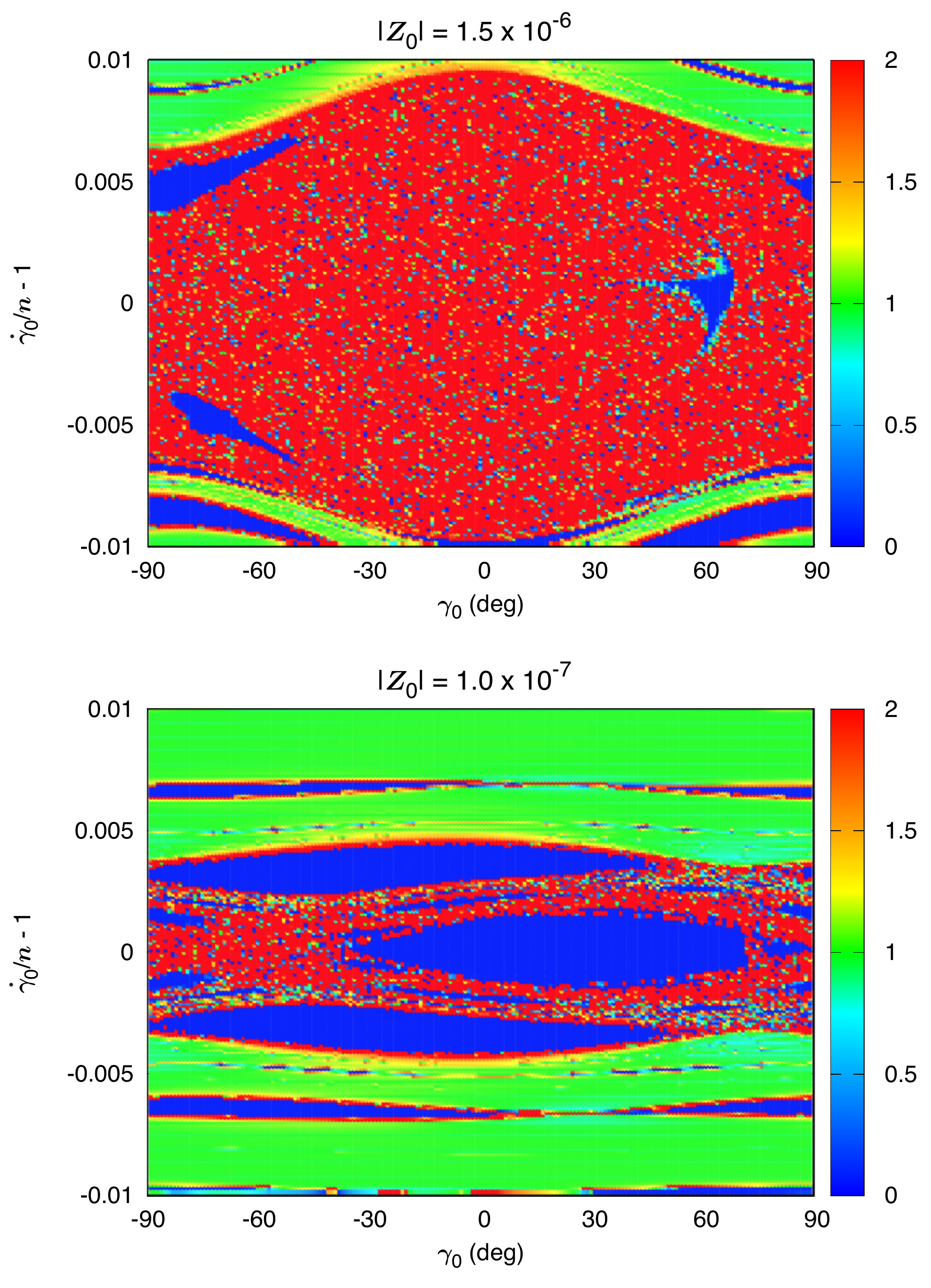}
 \caption{Frequency map analysis of KOI-1599 outer planet rotation angle assuming a rigid body with $|Z_0|=1.5 \times 10^{-6}$ (\textit{top}) or $|Z_0|=1.0\times 10^{-7}$ (\textit{bottom}). Each dot corresponds to a grid with $181 \times 201$ different initial conditions of $\gamma(0)$ and $\dot{\gamma}(0)$, respectively, with $\gamma(t) = \theta(t)-\lambda(0)$. The color index gives the derivative $\partial\eta/\partial\dot{\gamma}(0)$, where $\eta$ is the main frequency of $\gamma(t)$. The blue dots correspond to resonant motion, green dots to nonresonant regular motion, and red dots to chaotic motion.}
\label{SSKOI}
\end{figure}

In order to study in more detail the spin dynamics, we have to run direct numerical simulations of the rotation angle of the outer planet. 
We then integrated equation (\ref{130104d}) together with 
the orbital solution of the KOI-1599 system given by a classical $N-$body integrator.
In Figure~\ref{SSKOI} we show a frequency map analysis of the rotation angle \citep{Laskar_1993PD, Leleu_etal_2016} for the two $|Z_0|$ values in the neighborhood  of the synchronous resonance. 
We conclude that our predictions for the spin of the outer planet based on the \citet{Chirikov_1979} general criterion are correct: for $|Z_0| \sim 10^{-7}$ the planet can present stable asynchronous rotation, while for $|Z_0| \sim 10^{-6}$ it likely presents chaotic rotation.

\section{Tidal deformation and forced oscillations}
\llabel{forced}

Planets are not completely rigid bodies. 
They deform under tidal interactions with the parent star because the mass distribution inside the planet adjusts to the tidal potential \citep{Darwin_1879a, Darwin_1880}.
Viscoelastic rheologies have been shown to reproduce the main features of tidal effects.
The planet can respond as an elastic solid or as a viscous fluid, depending on the frequency of the perturbation \citep[for a review see][]{Henning_etal_2009}.

One of the simplest viscoelastic models is to consider that the planet behaves like a Maxwell material, which is represented by a purely viscous damper and a purely elastic spring connected in series \cite[e.g.,][]{Turcotte_Schubert_2002}.
More complex models, such as the Andrade model \citep{Andrade_1910}, can be thought of as the Maxwell model equipped with extra terms describing hereditary reaction of strain to stress \citep{Efroimsky_2012}. 
Therefore, the main features of tidal response are similar to the Maxwell model.

For simplicity, we adopt the Maxwell viscoelastic rheology.
The viscous response of the planet to the tidal excitation is modeled  by the parameter $\taub$, which corresponds to the relaxation time of the planet, while the elastic response in modeled  by the parameter $\taua$. 
The deformation $Z$ follows \citep{Correia_etal_2014} 
\be
  \label{eq:Z}
  Z + \taub \dot{Z} = Z_0 +
  Z_\mathrm{e} + \taua \dot{Z}_\mathrm{e} \ ,
\ee
with 
\be
  \label{eq:Ze}
  Z_\mathrm{e} =  \A \, \bar{x} \, \expo{2\im\theta} \ , 
  \quad \mathrm{where} \quad
\A = \frac{\kf}{4} \frac{\M}{\m} \left(\frac{R}{a}\right)^3 \ ,
\ee 
$\kf$ is the fluid second Love number for potential, and $\bar x$ is the complex conjugate of $x$. 
The function $Z_e$ is the equilibrium shape, i.e., the maximal deformation that the planet can attain owing to the tidal perturbation.

We denote the average rotation rate of the planet by $\omg = \langle \dot \theta \rangle$.
For simplicity, we assume that $\theta$ has a small forced amplitude of libration due to orbital forcing and a zero free amplitude of libration, such that 
\begin{eqnarray}
  \label{eq:theta}
  \theta &=& \omg t
  + \sum_{\vect{k}} \theta_\vect{k} \expo{\im \vect{k} \cdot \vect{\nu} t} \ , \\
 \label{eq:eitheta}
  \expo{2\im\theta} &=& \expo{2\im\omg t}
  \sum_{j\geq 0} \frac{\left(\im \sum_\vect{k} 2\theta_\vect{k} \expo{\im \vect{k} \cdot \vect{\nu} t}\right)^j}{j!} = \sum_{\vect{l}} y_\vect{l} \expo{\im (2 \omg + \vect{l} \cdot \vect{\nu}) t} \ ,
\end{eqnarray}
where all $\theta_\vect{k}$ are small, and the series is absolutely convergent ($\sum_\vect{k}|\theta_\vect{k}|$ converges).
Each coefficient $y_\vect{l}$ can be expressed as
a power series of the coefficients $\theta_\vect{k}$,
by identifying terms on both sides of Eq.~(\ref{eq:eitheta}).

From expressions (\ref{eq:x}), (\ref{eq:Ze}), and (\ref{eq:eitheta}), we obtain
\be
Z_e = \A \sum_\vect{k} \left(\sum_\vect{l} \bar x_\vect{k+l} y_\vect{l}\right) \expo{\im (2 \omg - \vect{k} \cdot \vect{\nu}) t} \ ,
\ee
and replacing this in expression (\ref{eq:Z}), we get
\be
 Z = Z_0  + Z_t + C \expo{-t/\taub} \ ,
 \llabel{190508a}
 \ee
 with
 \be
Z_t = \A \sum_\vect{k} \left(\sum_\vect{l} \bar x_\vect{k+l} y_\vect{l}\right) \frac{1+\im(2 \omg - \vect{k} \cdot \vect{\nu})\taua}{1+\im(2 \omg - \vect{k} \cdot \vect{\nu})\taub} \expo{\im (2 \omg - \vect{k} \cdot \vect{\nu}) t} 
\ . \llabel{190508b}
\ee
In the following, we neglect the term $C\expo{-t/\taub}$ because it vanishes after some time, and we only consider the solution $Z=Z_0+Z_t$.


\subsection{No resonant rotation}

We assume that $\omg \ne \nu_\vect{k} = \vect{k} \cdot \vect{\nu} / 2$ for all $\vect{k}$ (no resonant rotation).
Then, over one rotational period we have (Eq.\,(\ref{190508b}))
\be
\langle Z_t \rangle  = 0 \ ,
\ee
that is, there is no permanent tidal deformation.

\subsection{Resonant rotation without forced oscillations}
\label{sec:noforced}

The situation changes if the rotation is locked in some spin-orbit resonance, that is, we assume that $\omg = \nu_\vect{r} = \vect{r} \cdot \vect{\nu}/2$ (Eq.\,(\ref{190222a})).
We can then rewrite expression (\ref{190508a}) as
\be
Z = \sum_\vect{k} z_\vect{k} \expo{\im (\vect{r} - \vect{k}) \cdot \vect{\nu} t} \ ,
\llabel{190506b}
\ee
with
\be
  \label{eq:zk}
  z_\vect{k}  = Z_0 \, \delta_{\vect{r},\vect{k}} +
  \A \, \b_{\vect{r}-\vect{k}} \sum_\vect{l} \bar x_\vect{k+l} y_\vect{l} \ ,
\ee
and
\be
\b_{\vect{k}} = \frac{1+\im(\vect{k} \cdot \vect{\nu}) \taua}{1+\im(\vect{k} \cdot \vect{\nu}) \taub} 
\ .  \label{190526c}
\ee
The $\b_{\vect{k}}$ function depends on the tidal model, in this case, the Maxwell model.
Different tidal models would imply different expressions for $\b_{\vect{k}}$, but for all models we get $\b_{\vect{0}}=1$.

In the case of very weak or no forced oscillations,
expressions (\ref{eq:theta}) and (\ref{eq:eitheta}) simplify as
\begin{eqnarray}
  \label{eq:thetanoforce}
  \theta &=& \theta_0 + \frac{\vect{r} \cdot \vect{\nu}}{2} t \ , \\
  \label{eq:eithetanoforce}
  \expo{2\im\theta} &=& \expo{2 \im \theta_0} \expo{\im \vect{r} \cdot \vect{\nu} t} \ ,
\end{eqnarray}
that is, $y_\vect{0} = \expo{2\im\theta_0}$, and
$y_\vect{k \ne 0} = 0$.
Replacing in expression (\ref{eq:zk}), we get for the deformation of the planet
\be
  z_\vect{k}  = Z_0 \, \delta_{\vect{r},\vect{k}} +
  \A \, \b_{\vect{r}-\vect{k}}  \, \expo{2 \im \theta_0} \, \bar x_\vect{k} 
 \llabel{190520e} \ .
 \ee
Finally, to get the permanent tidal deformation, we can average over one rotational period 
\be
\langle Z_t \rangle = \langle Z \rangle - Z_0 = 
  \A \, \expo{2\im \theta_0} \, \bar x_{\vect{r}} \ ,
\llabel{190417b}
\ee
or
\be
|\langle Z_t \rangle | =   \A \, | x_{\vect{r}}| \ .
\llabel{190417c}
\ee

For instance, in the classical case of a single planet in an elliptical orbit (section~\ref{spc}), we have (Eq.\,(\ref{190417a}))
\be
\bar x_k = \Xh_k^{-3,2} (e) \, \expo{-2\im \varpi} \ ,
\ee
thus, for a planet trapped in the $\nu_k / n = k/2$ resonance
\be
| \langle Z_t \rangle | =  \A \, \Xh_k^{-3,2} (e) \ .
\ee

For the outer planet in the KOI-1599 system (section \ref{mps}), we have for the synchronous resonance (Table~\ref{TkoiAF})
\be
|x_{[2,0]}| \approx 0.93 \quad \Rightarrow \quad | \langle Z_t \rangle | \approx  2.5 \times 10^{-6} \ ,
\llabel{190527a}
\ee
while for the super- and sub-synchronous resonances
\be
|x_{[2,\pm1]}| \approx 0.25 \quad \Rightarrow \quad | \langle Z_t \rangle | \approx  7.0 \times 10^{-7} \ .
\llabel{190523z}
\ee

We note that $ \langle Z_t \rangle $ (Eq.\,(\ref{190417b})) does not depend on the tidal model function $\b_{\vect{k}}$ (Eq.\,(\ref{190526c})). 
Indeed, $ \langle Z_t \rangle $ is independent of the tidal model, as it corresponds to the maximal deformation that the planet can achieve while trapped in a given resonance.

\subsection{Resonant rotation with forced oscillations}
\label{sec:forced}

In case of resonant motion ($\omg = \nu_\vect{r}$), replacing the deformation of the planet, $Z$ (Eq.\,(\ref{190506b})), in expression (\ref{eq:ddtheta}), we get
\be
  \ddot{\theta} 
  = \B \, \Im\left(\sum_\vect{k}\left(
  \sum_{\vect{l},\vect{m}} x_\vect{k+l+m} \bar y_\vect{l} z_\vect{m}
  \right) \expo{\im \vect{k} \cdot \vect{\nu} t} \right) \ ,
  \llabel{190507a}
\ee
and from expression (\ref{eq:theta})
\be
  \ddot{\theta} = - \sum_\vect{k} (\vect{k} \cdot \vect{\nu})^2 \theta_\vect{k} \expo{\im \vect{k} \cdot \vect{\nu} t} \ .
\ee
Then, by identifying terms in both series, we find
\be
  \label{eq:thetak}
  \theta_\vect{k} = \frac{1}{2\im} \left( \alpha_\vect{k} - \bar \alpha_\vect{-k} \right) \ ,
\ee
where
\be
\alpha_\vect{k} = \frac{\B}{(\vect{k} \cdot \vect{\nu})^2} \sum_{\vect{l},\vect{m}} \bar{x}_\vect{l+m-k} y_\vect{l} \bar z_\vect{m}
\ . \llabel{190520b}
\ee
We also note that $\theta_{-\vect{k}} = \bar \theta_\vect{k}$. 
Therefore, with $\phi_\vect{k} = \arg \alpha_\vect{k}$, we can rewrite expression (\ref{eq:theta}) as 
\be
  \theta = \nu_\vect{r} \, t + \theta_0 + \sum_{\vect{k} \ne 0} | \alpha_\vect{k} | \sin (\vect{k} \cdot \vect{\nu} \, t + \phi_\vect{k}) 
\ . \llabel{190520a}
\ee

We recall that the coefficients $y_\vect{l}$ are expressed as power series of the coefficients $\theta_\vect{k}$ (Eq.\,(\ref{eq:eitheta})), and that the coefficients $z_\vect{k}$ are linear combinations of the coefficients $y_\vect{l}$ (Eq.\,(\ref{eq:zk})).
Equations~(\ref{eq:eitheta}), (\ref{eq:zk}), and (\ref{eq:thetak}) thus define a closed system of equations, which can be solved for the amplitudes of the forced oscillations $\theta_\vect{k}$ and for the deformation of the planet $z_\vect{k}$.

We can estimate the order of magnitude of the expected forced oscillations.
Since $|x_\vect{k}| \le 1$, $|y_\vect{k}| \le 1$ and $|z_\vect{k}| \le \Z = \max (|Z_0|, \A)$,
we deduce from expression (\ref{eq:thetak})
\be
|\theta_\vect{k}| \le \frac{\B \, \Z}{ (\vect{k}\cdot\vect{\nu})^2} \approx 18 \, \Z \left(\frac{n}{2 \nu_\vect{k}}\right)^2 \ .
\ee
For an Earth-size planet and a perturbing frequency on the order of the orbital frequency
($\nu_\vect{k} \sim n$),
we get $|\theta_\vect{k}| \lesssim 10^{-5}$, that is,
the amplitude of the forced oscillations remains very small.
We thus linearize expression~(\ref{eq:eitheta}) and find
\be
y_\vect{0} = \expo{\im 2 \theta_0} \ , \quad \mathrm{and} \quad
y_\vect{l \ne 0} \approx  2 \im \, \theta_\vect{l} \, \expo{\im 2 \theta_0} = \left( \alpha_\vect{l} - \bar \alpha_\vect{-l} \right) \expo{\im 2 \theta_0}
\llabel{190520d} \ .
\ee
The coefficients $z_\vect{k}$ thus become linear in $\theta_\vect{k}$,
so we only need to solve a system of linear equations to get the amplitudes of forced libration,
$\theta_\vect{k}$, and the corresponding corrections to the deformation of the planet, $z_\vect{k}$.
For planets involved in mean-motion resonances, we may have $\nu_\vect{k}/n\sim 0.01$ \citep{Delisle_etal_2017}, so the amplitude of forced oscillations is expected to be stronger ($|\theta_\vect{k}| \lesssim 0.1$).
In this case, higher order terms should be included and we would need to solve a system of polynomial equations
\citep[see][Appendix~D]{Delisle_etal_2017}.

For the sake of simplicity and readability,
we restrict our study to the first order case in the following,
as it already provides a good approximation in most cases.
Then, from expressions (\ref{190520b}) and (\ref{190520d}) we have
\be
\alpha_\vect{k} \approx \frac{18 \, n^2 \, \expo{\im 2 \theta_0} }{(\vect{k} \cdot \vect{\nu})^2} \sum_\vect{m} \bar{x}_\vect{m-k} \bar z_\vect{m} 
\ , \llabel{190520c} 
\ee
where $\bar z_\vect{m}$ is given by expression (\ref{eq:zk}).
In order to be captured in a given resonance ($\omg = \nu_\vect{r}$), we have that $x_\vect{r} \ne 0$ (section~\ref{rigid}).
Since we always have $\vect{k} \ne \vect{0}$ (Eq.\,(\ref{190520a})),
it is needed that at least another $x_\vect{m \ne r} \ne 0$ to observe forced librations.
Indeed, if there is only one orbital forcing term, as in the case of a single planet in a circular orbit, all $\alpha_\vect{k}=0$, and hence there are no forced librations.

\subsubsection{Single planet case}
\label{spcforced}

We first consider a single planet in an elliptical orbit with small eccentricity.
The two dominating terms of the orbital forcing are $x_2 \approx 1$ and $x_3 \approx 7e/2$ (section~\ref{spc}), so we neglect the remaining terms.
Applying to the case of Mercury, we have $r=3$, $\nu=n$, and $|Z_0| \gg \A$.
Thus, expression (\ref{190520c}) becomes
\be
\alpha_k \approx \frac{18 \, \expo{\im 2 \theta_0} }{k^2} \, \bar Z_0 \, \bar{x}_{3-k}  
\ . \llabel{190526a} 
\ee
The only $\alpha_k \ne 0$ is for $x_{3-k} = x_2$, that is, for $k=1$.
Then, the rotation angle becomes (Eq.\,(\ref{190520a}))
\be
  \theta \approx \frac32 n t + \theta_0 + 18 \, | Z_0 \, x_2 | \sin (n t + \phi_1) 
\ . \llabel{190520f}
\ee

This simple approximation gives a forced libration period equal to the orbital period, and a forced libration amplitude $|\alpha_1| \approx 18 \, |Z_0| = 56''$, in perfect agreement with a full numerical simulation for Mercury \citep[see Fig.~2 in][]{Peale_etal_2007}.

\subsubsection{Multi-planet systems}

We consider the outer planet in the KOI-1599 system, for which the dominating terms of the orbital forcing are $x_{[2,0]} = 0.93$, $x_{[2,1]} = 0.26$, and $x_{[2,-1]} = 0.25$ (Table~\ref{TkoiAF}), and we 
neglect the remaining terms.
In this case, we have $|Z_0| \ll \A$, so replacing expression (\ref{eq:zk}) in (\ref{190520c}) gives
\be
\alpha_\vect{k} \approx \frac{18 \, \A \, n^2 \, \expo{\im 2 \theta_0} }{(\vect{k} \cdot \vect{\nu})^2} \sum_\vect{m} \b_{\vect{m}-\vect{r}} \, \bar{x}_\vect{m-k} \sum_\vect{l}  x_\vect{m+l} \, \bar y_\vect{l} 
\ , \llabel{190526b} 
\ee
which depends on the tidal model through $ \b_{\vect{m}-\vect{r}}$ (Eq.\,(\ref{190526c})).
For simplicity, we further assume that $\taua = 0$ and $\taub \nu_\vect{k} \gg 1$ (high frequency regime, see section~\ref{capture}), such that
$\b_{\vect{m}-\vect{r}} \approx \delta_{\vect{r},\vect{m}}$.
Then
\be
\alpha_\vect{k} \approx \frac{18 \, \A \, n^2 \, \expo{\im 2 \theta_0} }{(\vect{k} \cdot \vect{\nu})^2} \,
\bar{x}_\vect{r-k} \sum_\vect{l}  x_\vect{r+l} \, \bar y_\vect{l} 
\ . \llabel{190526z} 
\ee

\begin{figure}
\centering
    \includegraphics[width=\columnwidth]{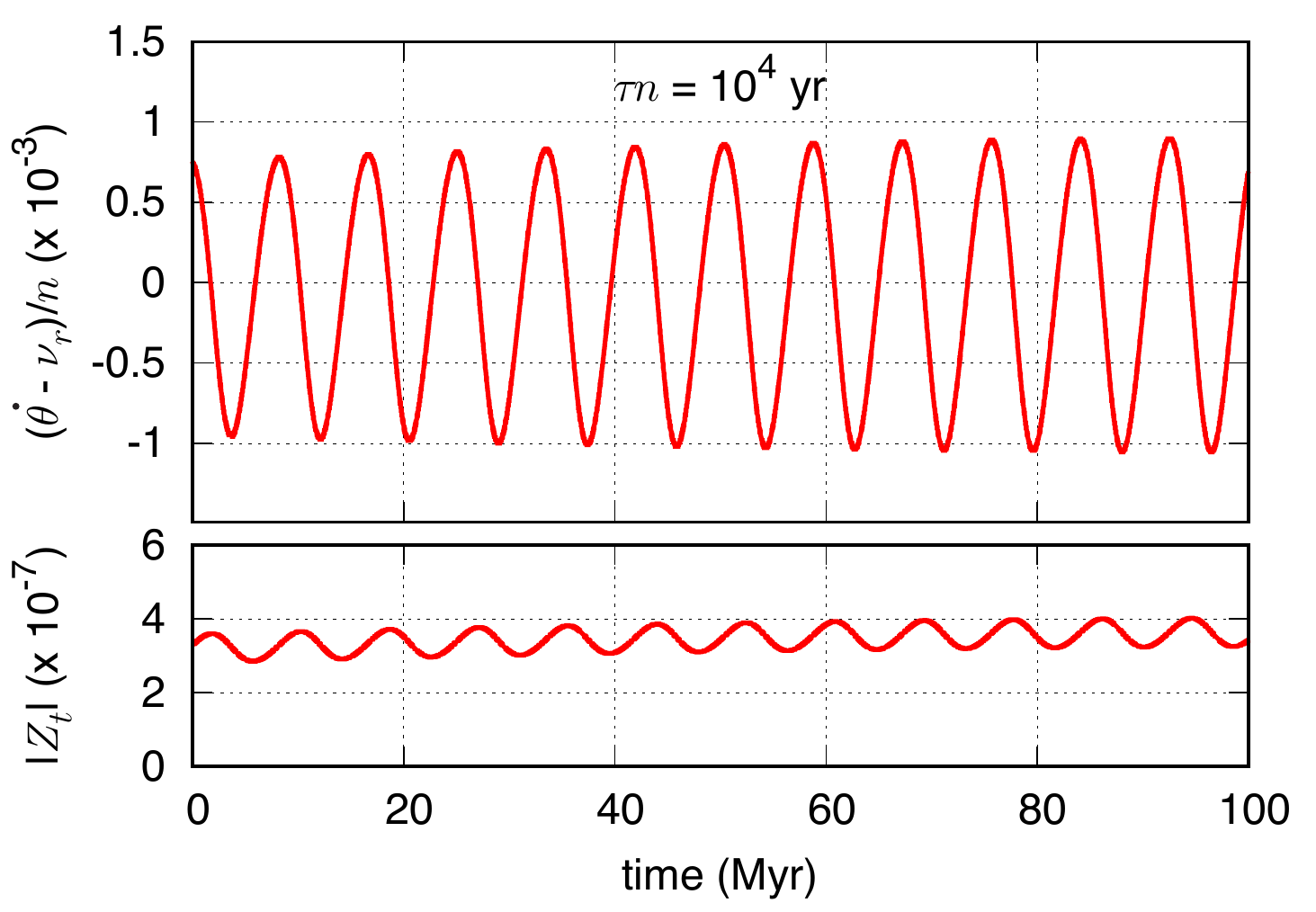}
 \caption{Forced libration of the rotation rate for the super-synchronous resonance ($\nu_\vect{r} = n + \lbf/2$) of the outer planet in the KOI-1599 system. We show the evolution of $(\dot \theta-\nu_\vect{r})/n$ adopting $\taua=0$ and $\taub n = 10^4$.} 
\label{libkoi}
\end{figure}

Let us consider, for example, that the planet is captured in the $\vect{r} = [2,1]$ spin-orbit resonance, with frequency $\nu_{[2,1]} = n + \lbf / 2$.
The only two $\alpha_\vect{k} \ne 0$ are for $\bar x_{[2,1]-\vect{k}} = \bar x_{[2,0]}$ and $\bar x_{[2,-1]}$, that is, for $\vect{k}=[0,1]$ and $[0,2]$, respectively, with associated frequencies $\lbf$ and $2 \lbf$, where $\lbf$ is the orbital libration frequency
\be
\alpha_{\lbf} \approx 18 \, \A \, \expo{\im 2 \theta_0} \left( \frac{n}{\lbf} \right)^2 \bar{x}_{[2,0]} \sum_\vect{l}  x_{[2,1]+\vect{l}} \, \bar y_\vect{l}   
\ , \llabel{190618a}
\ee
\be
\alpha_{2\lbf} \approx 18 \, \A \, \expo{\im 2 \theta_0} \left( \frac{n}{2 \lbf} \right)^2 \bar{x}_{[2,-1]}  \sum_\vect{l}  x_{[2,1]+\vect{l}} \, \bar y_\vect{l}  
\ . \llabel{190618b}
\ee
Since $| \alpha_{2\lbf} / \alpha_{\lbf} | \approx 0.067$, we can also neglect the contribution of $\alpha_{2\lbf}$. 
Thus, the rotation angle (Eq.\,(\ref{190520a})) simply becomes
\be
  \theta \approx (n+\tfrac12 \lbf) \, t + \theta_0 + |\alpha_{\lbf}| \sin (\lbf t + \phi_{\lbf}) 
\ , \llabel{190523b}
\ee
and the rotation rate
\be
\dot \theta \approx n + \lbf \left[ \tfrac{1}{2} + |\alpha_{\lbf}| \cos (\lbf t + \phi_{\lbf}) \right]
\ . \llabel{190523x}
\ee
From expression (\ref{190520d}) we additionally have
\begin{eqnarray}
\sum_\vect{l}  x_{[2,1]+\vect{l}} \, \bar y_\vect{l} &=&
 x_{[2,1]} \bar y_{[0,0]} +  x_{[2,0]} \bar y_{[0,-1]}  +  x_{[2,-1]} \bar y_{[0,-2]}  
\nonumber \\ &\approx&
\expo{-2\im \theta_0} \left( x_{[2,1]} -  x_{[2,0]} \, \alpha_{\lbf} \right)
\ . \llabel{190618c}
\end{eqnarray}
Inserting in expression (\ref{190618a}) and solving for $\alpha_{\lbf}$ we finally get
\be
\alpha_{\lbf} \approx \frac{18 \, \A  \left( n/\lbf \right)^2 \bar{x}_{[2,0]}  x_{[2,1]}}{1+18 \, \A  \left( n/\lbf \right)^2 |x_{[2,0]}|^2}
\ . \llabel{190618d}
\ee
With $\A = 2.7 \times 10^{-6}$ (Eq.\,(\ref{eq:Ze})), and $(n/\lbf) \approx 150$ (Table~\ref{TkoiAF}), we estimate $|\alpha_{\lbf}| \approx 0.14$~rad.
From expression (\ref{190523x}) we also estimate $\Delta \dot \theta / n = (\lbf/n) |\alpha_{\lbf} | \approx 9 \times 10^{-4}$.
In Figure~\ref{libkoi} we show the forced librations of the rotation rate for the $\vect{r} = [2,1]$ spin-orbit resonance, which is obtained performing a numerical simulation of the KOI-1599 system with $\taub n = 10^4$ (that is, $\taub \lbf \sim 10^2$).
We observe that the libration frequency of the spin is indeed $\lbf$, 
and that the forced libration amplitude also matches the estimated value. 

We can now estimate the correction that the forced libration introduces in the tidal deformation (Eq.\,(\ref{eq:zk})) as follows:
\begin{eqnarray}
\langle Z_t \rangle &\approx& 
\A \left( \bar x_{[2,1]} y_{[0,0]} + \bar x_{[2,0]} y_{[0,-1]} \right) 
\nonumber \\ &=&
\A \, \expo{2\im \theta_0} \left( \bar x_{[2,1]}  - \bar x_{[2,0]} \bar \alpha_{\lbf} \right)
\nonumber \\ &=&
\A \, \expo{2\im \theta_0} \bar x_{[2,1]} \left( 1  -  \frac{18 \, \A  \left( n/\lbf \right)^2  |x_{[2,0]}|^2}{1+18 \, \A  \left( n/\lbf \right)^2 |x_{[2,0]}|^2} \right) \ .
\llabel{190523c}
\end{eqnarray}
We thus obtain for the outer planet in the KOI-1599 system,
$|\langle Z_t \rangle | \approx 3.6 \times 10^{-7}$, which is about half of the value without considering the forced librations (Eq.\,(\ref{190523z})); this value corresponds to what is observed in the numerical simulations (Fig.\,\ref{libkoi}). 
We conclude that the forced libration contributes to decreasing the tidal deformation \citep[see also][Appendix D]{Delisle_etal_2017}.

\section{Tidal dissipation and capture probabilities}
\llabel{capture}

The tidal deformation of the planet is not instantaneous;
there is a delay between the perturbation and maximal deformation that leads to a nonzero net torque on the spin of the planet.
As a consequence, tidal effects slowly modify the average rotation rate of the planet
and may drive it into the different configurations described in section~\ref{rigid}.

\subsection{Tidal dissipation}

Replacing the tidal deformation of the planet $Z_t$ (Eq.\,(\ref{190508b})) in the general expression of the torque (Eq.\,(\ref{eq:ddtheta})), and averaging over the fundamental frequencies of the system we get
\begin{eqnarray}
\T 
&=& \left\langle \B \, \Im \left(Z_t \,x \, \expo{-2\im\theta}\right) \right\rangle
\nonumber \\  &=&  
\A \B \, \Im \left( \sum_{\vect{k},\vect{l},\vect{m}} x_\vect{k+l} \bar x_\vect{k+m} y_\vect{m} \bar y_\vect{l} \frac{1+\im(2 \omg - \vect{k} \cdot \vect{\nu})\taua}{1+\im(2 \omg - \vect{k} \cdot \vect{\nu})\taub} \right)
   \ . \llabel{190507b}
\end{eqnarray}
In addition, assuming very small or no forced oscillations, that is, $\theta \approx \langle \theta \rangle = \theta_0 + \omg \, t$ (Eq.\,(\ref{eq:theta})), we have $y_\vect{0} = \expo{2\im\theta_0}$ and
$y_\vect{k \ne 0} = 0$, so the average net torque can be simplified as
\be 
\T (\omg) =  - \K \, \sum_{\vect{k}} | x_\vect{k} |^2  \frac{(2 \omg - \vect{k} \cdot \vect{\nu})\taub}{1+(2 \omg - \vect{k} \cdot \vect{\nu})^2 \taub^2}  
\ , \llabel{190507c}
\ee
where 
\be
\K = \A \B \left(1 - \frac{\taua}{\taub} \right) = \kf \frac{3 G \M^2 R^5}{2 C a^6} \left( 1 - \frac{\taua}{\taub} \right) \ .
\ee

The initial rotation rate of the planets in unknown.
A small number of large impacts at the end of the formation process can significantly change the spin rate or direction \citep{Dones_Tremaine_1993, Kokubo_Ida_2007}.
However, since planets contract from an initial larger protoplanet, in general it is expected that they rotate fast.
Indeed, based on the solar system planets, \citet{MacDonald_1964} proposed an empirical relation $\omg_0 \propto m^{4/5} / R^2$, which leads to about 12~h for Earth-size planets. 
Therefore, in general we expect $\omg_0 \gg \vect{k} \cdot \vect{\nu}$, and since $\taua<\taub$, the initial rotation rate can only decrease (Eq.\,(\ref{190507c})).

\begin{figure*}
\centering
    \includegraphics[width=\textwidth,height=\textheight*2/3]{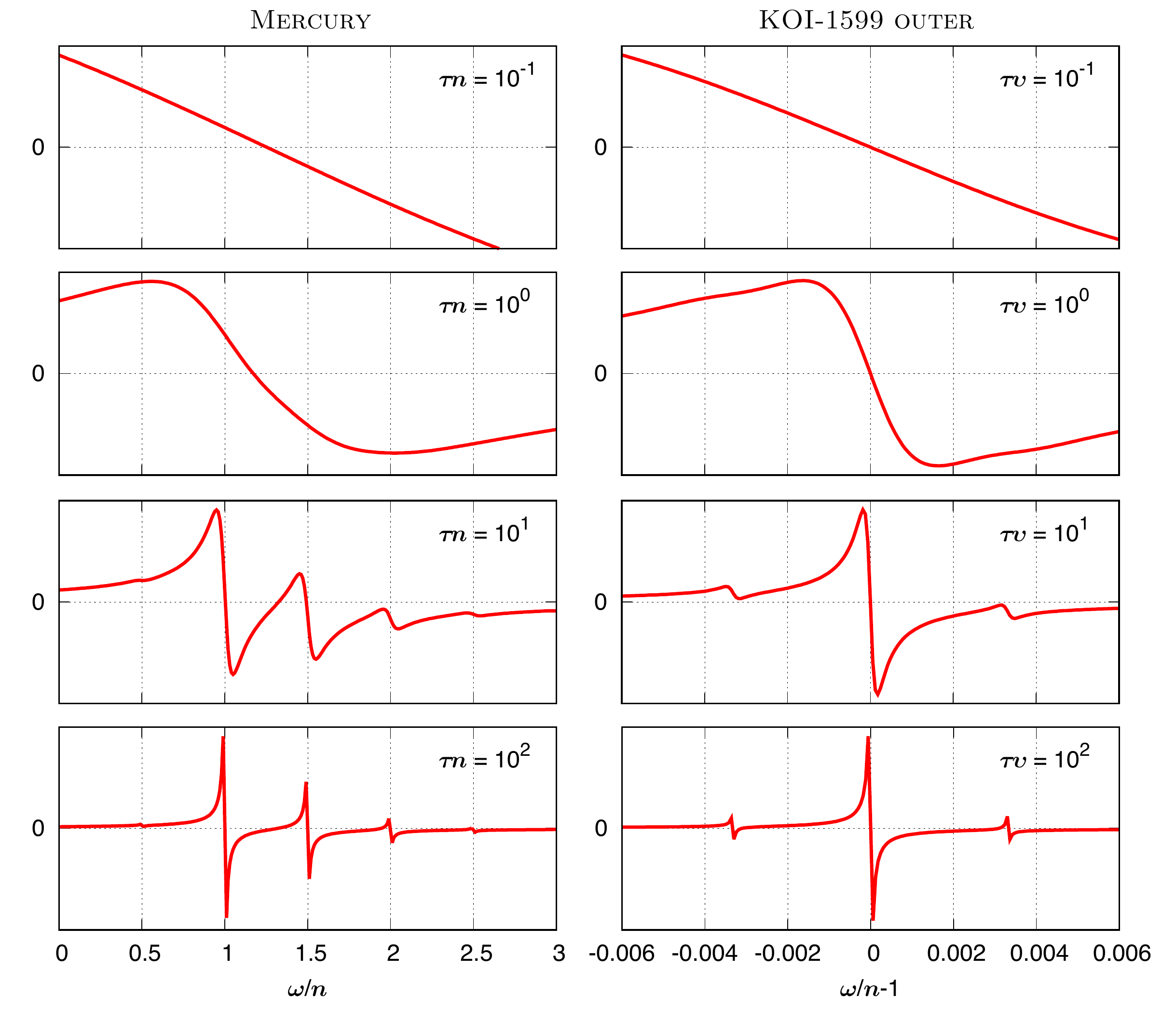}
 \caption{Normalized net torque $\T / \K$ (Eq.\,(\ref{190507c})) for different $\taub$ values and two orbital forcing functions: Mercury, for which $|x_k| = \Xh_k^{-3,2}(e)$ and $e=0.206$ (left column); the outer planet in the KOI-1599 system, for which the $|x_k|$ are given in Table~\ref{TkoiAF} (right column). } 
\label{NetTorque}
\end{figure*}

As the planet arrives in a regime where $\omg \sim \vect{k} \cdot \vect{\nu}$ the situation changes.
Depending on $\taub$ and the amplitudes $|x_k|$, the torque may become positive, and hence create stable equilibria for the rotation rate.
In Figure~\ref{NetTorque} we draw the net torque as a function of $\taub$ for two different orbital forcing functions.
The left-hand column represents the case of Mercury, for which $|x_k| = \Xh_k^{-3,2}(e)$ with $e=0.206$, while the right-hand column is for the outer planet in the KOI-1599 system, for which $|x_k|$ are given in Table~\ref{TkoiAF}.

We observe that, for small $\taub$ values, only one equilibrium point near the synchronous rotation is possible.
However, for large $\taub$ values, several equilibria are possible, and they coincide exactly with the resonant islands identified in section~\ref{rigid}.
This result was already known for the classical case of a single planet in an eccentric orbit \citep[see][]{Correia_etal_2014}, but it is remarkable that it prevails for any kind of orbital forcing.
This property of the tidal torque enhances the chances of observing the rotation of close-in planets trapped in some asynchronous resonance.

\subsection{Capture probabilities}

The total torque on the rotation (Eq.\,(\ref{eq:ddtheta})) can be written as
\be
\ddot{\theta} = \B \, \Im \left(Z_0 \,x \, \expo{-2\im\theta}\right) + \B \, \Im \left(Z_t \,x \, \expo{-2\im\theta}\right)
\ , \label{190508c}
\ee
since $Z=Z_0+Z_t$.
The average of the second term is the tidal torque $\T (\omg)$ (Eq.\,(\ref{190507c})).
The average of the first term is different from zero when the rotation rate is close to some resonant island,
$\omg \approx \nu_\vect{r}$, with $\nu_\vect{r} = \vect{r}\cdot\vect{\nu}/2$ (Eq.\,(\ref{eq:ddtheta0})):
\be
\left\langle \B \, \Im \left(Z_0 \,x \, \expo{-2\im\theta}\right) \right\rangle = 
 \B \, \Im \left(Z_0  x_\vect{r} \, \expo{- 2 \im (\gam-\phi)} \right) \ ,
\ee
where $\gam = \theta - \nu_\vect{r} \, t + \phi$, and $\phi = \tfrac12 \arg (Z_0 x_r)$ is a constant phase.
Thus, the average of the total torque (Eq.\,(\ref{190508c})) becomes 
\be
\ddot{\gam} = - \frac{\beta_\vect{r}^2}{2} \sin 2 \gam + \T(\nu_\vect{r}+\dot \gam)
\ , \label{021014g}
\ee
where $\beta_\vect{r}^2 = 2 \B |  Z_0 x_\vect{r} |$ (Eq.\,(\ref{190225a})).
If $Z_0=0$, capture could only be possible at the equilibria of the tidal torque, that is, when $\T=0$.
However, in a more general situation, the tidal torque can be counterbalanced by the conservative torque resulting from $\beta_\vect{r}$.
Therefore, even when $\T \ne 0$, each time the planet crosses a resonant island, there is some chance of being captured.

\citet{Goldreich_Peale_1966} computed a first estimation of 
the capture probability $ P_\mathrm{cap} $,  
and subsequently more detailed studies proved their expression to be
essentially correct \citep[for a review, see][]{Henrard_1993}.
We follow the same approach described in section~5.3 of \citet{Correia_Laskar_2010}.
Expression (\ref{021014g}) can be rewritten as
\be
 \frac{d h}{d t} = 2 \, \T  \, \frac{d \gam}{d t} \ ,
\ee 
where
\be 
h = \dot \gam^2 -  \frac{\beta_\vect{r}^2}{2} \cos 2 \gam 
\llabel{190508y}
\ee
is an integral of motion in the conservative problem (section~\ref{rigid}), which can be related to the total energy.
The separatrix equation between dynamical regimes is $h = \beta_\vect{r}^2/2$, where $h > \beta_\vect{r}^2/2$ gives the trajectories in the circulation zone (outside the resonance) and $h < \beta_\vect{r}^2/2$ the trajectories in the libration zone (captured in resonance). 
The ``energy'' variation of the planet after a cycle around the
resonance is then given by the function
\be
\Delta h ( \gam_1 ) =2 \int_{\gam_1}^{\gam_2} \T  \, d \gam + 2
\int_{\gam_2}^{\gam_3} \T  \, d \gam \ , \nonumber
\ee
where $ \gam_1 $ is the $ \gam $ value when the planet crosses the
separatrix between the circulation and libration zones.
The values $ \gam_2 $ and $ \gam_3 $ are the first two following $ \gam $ values
corresponding to $ \dot \gam = 0 $.
In order to be captured, 
after a cycle inside the resonance, the planet must remain within the libration zone, that is, the ``energy'' variation after a cycle must be negative $\Delta h ( \gam_1 ) < 0$.
Assuming that $ \gam_1 $ is uniformly distributed 
between $ - \pi / 2 $ and $ \pi / 2 $, capture inside the resonance occurs for
``energy'' variations within
\be
\left\{ {\cal E} : \Delta h ( - \pi / 2 ) < \Delta h ( \gam_1 ) < 0 \right\} \ ,
\ee
from a total of possibilities
\be
\left\{ {\cal E}_T : \Delta h ( - \pi / 2 ) < \Delta h ( \gam_1 ) < \Delta h (
\pi / 2 ) \right\} \ ,
\ee
that is,
\be
P_\mathrm{cap} = \frac{\int_{\cal E} d \gam_1}{\int_{{\cal E}_T} d \gam_1} =
\frac{\Delta h ( - \frac{\pi}{2} )}{\Delta h ( - \frac{\pi}{2} ) - \Delta h (
\frac{\pi}{2} )} \llabel{021024k} \ .
\ee

Usually the tidal torque is very weak, that is, $ |\T (\dot \gam) |
\ll \beta_\vect{r}^2$, which gives $ \gam_3 \simeq - \pi / 2 $, $ \gam_2 \simeq \pi / 2 $, and $h \approx  \beta_\vect{r}^2/2$.
Then, from expression (\ref{190508y}) we have
\be 
\dot \gam = \mathrm{sign} ( \dot \gam )\,  \beta_\vect{r} \cos \gam \ ,
\llabel{090113a}
\ee
and replacing it in expression (\ref{021024k}) we finally get
\be
 P_\mathrm{cap} = 1 - \frac{T_-}{T_+} \ , \quad \mathrm{with} \quad
 T_\pm = \int_{-\frac{\pi}{2}}^{\frac{\pi}{2}} \T \left( \nu_\vect{r}  \pm \beta_\vect{r} \cos \gam \right) d \gam 
 \ . \llabel{021024d}
 \ee
This expression is valid for any kind of tidal torque.
For the Maxwell rheology that we adopt in this paper, we use expression (\ref{190507c}) for $\T$, which leads to
\begin{eqnarray}
T_\pm = - \K \, \sum_\vect{k} | x_\vect{k} |^2 \left[ \frac{1}{\Delta_\vect{k}^+} \ln \left( \frac{1+\delta_\vect{k}^+}{1-\delta_\vect{k}^+} \right) - \frac{1}{\Delta_\vect{k}^-} \ln \left( \frac{1+\delta_\vect{k}^-}{1-\delta_\vect{k}^-} \right) \right] \ ,
\llabel{190627a}
\end{eqnarray}
with
$\Delta_\vect{k}^\pm = \sqrt{\eta^2-(\lambda_\vect{k} \pm \im)^2}$, $\delta_\vect{k}^\pm = (\pm(\eta-\lambda_\vect{k})-\im)/\Delta_\vect{k}^\pm$, $\eta = \pm 2 \taub \beta_\vect{r}$, and $\lambda_\vect{k} = \taub (\vect{r} - \vect{k}) \cdot \vect{\nu}$.
The $\pm$ in $\eta$ corresponds to $T_\pm$.
Since both $\eta$ and $\lambda_\vect{k}$ depend on $\taub$ and $|Z_0|$ (through $\beta_\vect{r}$), the chances of capture in each resonant island change with these two parameters.

\begin{figure}
\centering
    \includegraphics[width=\columnwidth]{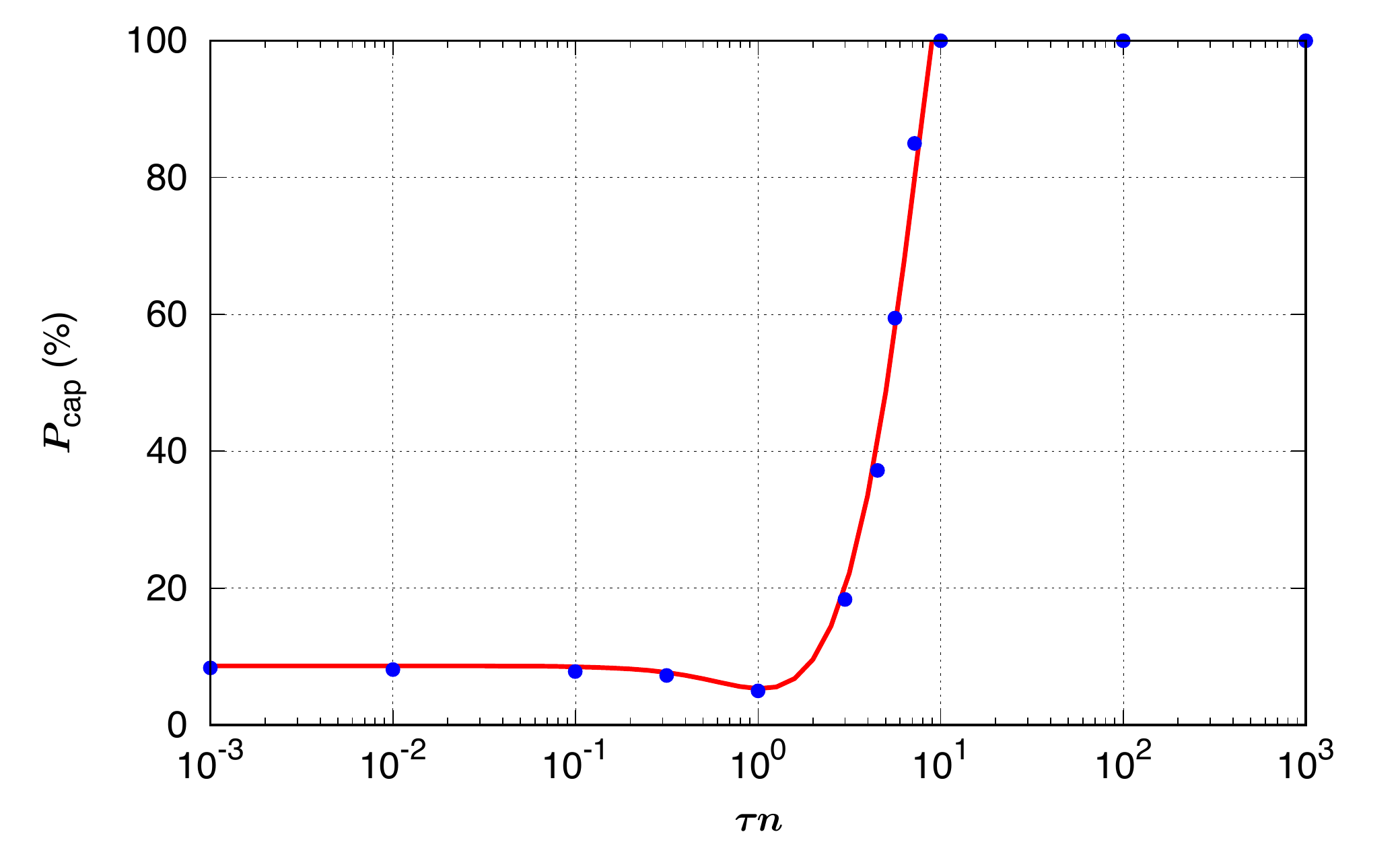}
 \caption{Capture probabilities in the presently observed 3/2 spin-orbit resonance for Mercury ($e=0.206$ and $|Z_0| = 1.25 \times 10^{-5}$). The solid line gives the theoretical approximation given by expression (\ref{021024d}), while the dots give the results of numerical simulations with equations (\ref{130104d}) and (\ref{eq:Z}). We run 180 initial conditions with $\theta_0$ differing by one degree.}
\label{PcapM}
\end{figure}

In Figure~\ref{PcapM}, we show the capture probabilities in the presently observed 3/2 spin-orbit resonance for Mercury ($e=0.206$ and $|Z_0| = 1.25 \times 10^{-5}$).
We see there is a very good agreement between the theoretical approximation (Eq.\,(\ref{021024d})) and the results of numerical simulations (Eqs.\,(\ref{130104d}) and (\ref{eq:Z})).
For $\taub n < 0.1$, the probability is constant and equal to about 8.5\%, the same value obtained with a linear tidal model \citep{Correia_Laskar_2004}.
For $\taub n \sim 1$, the probability slightly decreases and then increases very fast.
For $\taub n > 10$, the probability is always 100\%.
The transition occurs because for $\taub n \sim 1$ the tidal torque changes from a regime proportional to the forcing frequency to a regime inversely proportional (Eq.\,(\ref{190507c})).
As a result, for $\taub n > 1$, multiple equilibria for the tidal torque are possible (Fig.\,\ref{NetTorque}).
One of this equilibrium points is placed exactly at $\omg/n = 3/2$, so the rotation rate of the planet is naturally driven to the corresponding resonant island and capture is unavoidable.
Indeed, for $\taub n > 10$, the planet can remain in the 3/2 spin-orbit equilibria even with $Z_0 = 0$ \citep{Correia_etal_2014}.

\begin{figure}
\centering
    \includegraphics[width=\columnwidth]{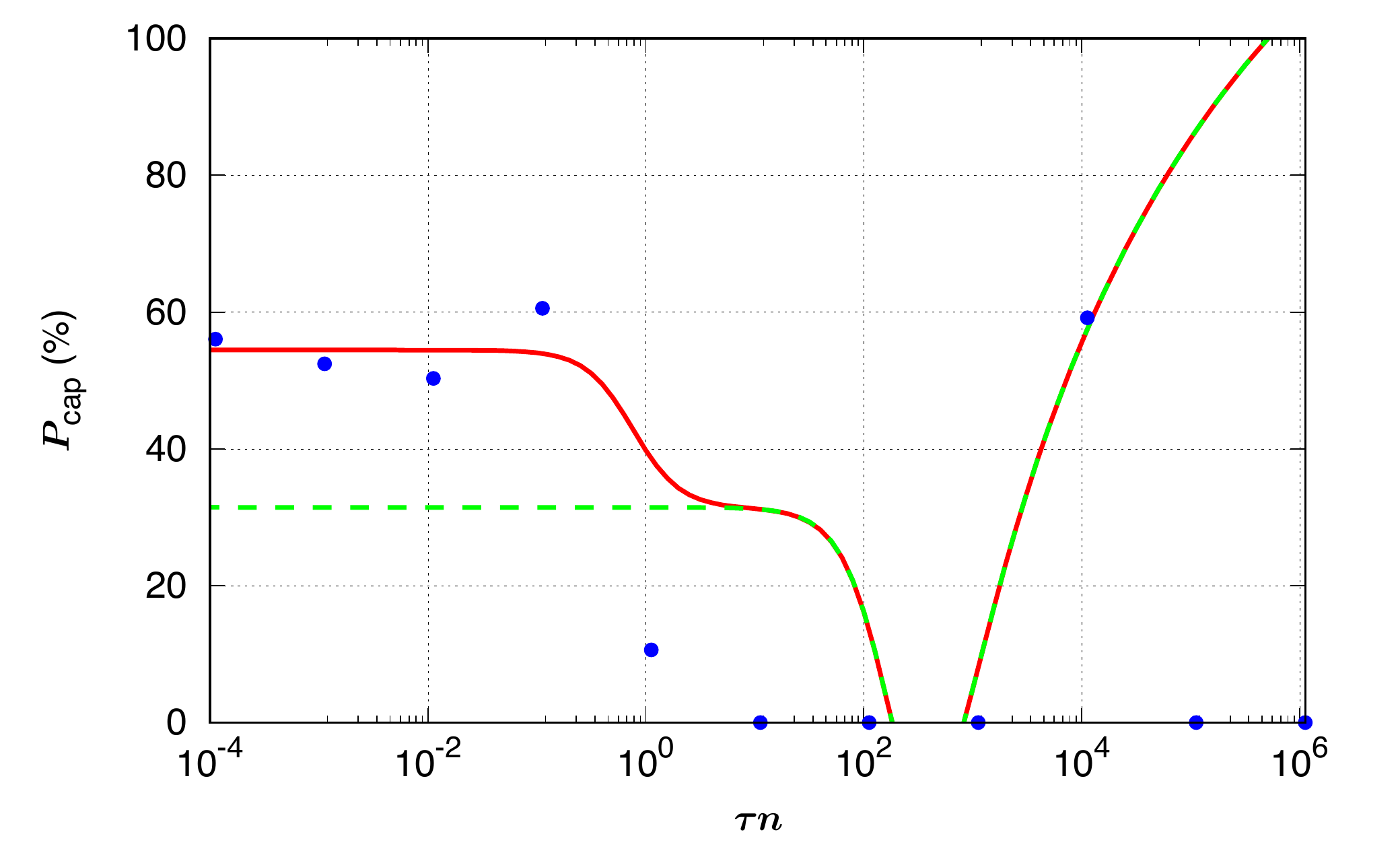}
 \caption{Capture probabilities for the super-synchronous resonance $\nu_\vect{r} = n + \nu/2$ for the outer planet in the KOI-1599 system with $|Z_0| = 10^{-7}$. The solid line gives the theoretical approximation given by expression (\ref{021024d}) with all terms from Table~\ref{TkoiAF}, while the dashed line is obtained with only the first three dominating terms. The dots give the results of numerical simulations with equations (\ref{130104d}) and (\ref{eq:Z}). We run 180 initial conditions with $\theta_0$ differing by one degree.}
\label{PcapKOI}
\end{figure}

In Figure~\ref{PcapKOI}, we show the theoretical capture probabilities (Eq.\,(\ref{021024d})) for the super-synchronous resonance $\nu_\vect{r} = n + \nu/2$ for the outer planet in the KOI-1599 system ($\vect{r} = [2,1]$, Table~\ref{TkoiAF}).
We adopt $|Z_0| = 10^{-7}$, since for larger values of $Z_0$ the libration widths of the individual resonances overlap, and thus it is no longer possible to capture the rotation in individual resonances (see Fig.\,\ref{FFTKOI}).
As for the case of Mercury, we observe that for $\taub n \ll 1$ the probability stabilizes at a constant value (54\%), while for $\taub n \gg 1$ the probability is 100\%.
However, there are two transitions of regime.
The main transition occurs around $\taub n \sim 10^2 \Leftrightarrow \taub \lbf \sim 1$ because the dominating terms of the orbital forcing are those involving the orbital libration frequency, $\lbf$ (Table~\ref{TkoiAF}).
If we only consider the main three terms of the orbital forcing, this transition, for which the capture probability drops to zero, would be the only observed (green dashed line).
The other transition occurs around $\taub n \sim 1$ and results from terms that involve the precession rate of the orbit, $g$. 
These terms have smaller amplitudes than in the case of Mercury because $e \approx 0.02$ (Table~\ref{TkoiGEAI}), but they contribute to increase the capture probabilities in the super-synchronous resonance.

In Figure~\ref{PcapKOI}, we additionally show the results from numerical simulations (Eqs.\,(\ref{130104d}) and (\ref{eq:Z})).
Contrary to the case of Mercury (Fig.\,\ref{PcapM}), in this case the agreement between the theoretical and numerical capture probabilities is not so good.
For $\taub n \ll 10^2$, this is because the theoretical probabilities (Eq.\,(\ref{021024d})) were estimated using the approximation (\ref{021014g}), which assumes that the effect of the nearby resonances can be averaged to zero.
However, although isolated resonant islands exist for $|Z_0| = 10^{-7}$, there is already a small chaotic zone around these islands (Fig.\,\ref{SSKOI}, bottom).
Therefore, in the numerical simulations the rotation is chaotic for some time before it is stabilized in one of these islands (see section~\ref{evol}) and the probabilities are altered.
For $\taub n \gg 10^2$, in addition to the previous effect, there is a significant increase in the permanent tidal deformation $\langle Z_t \rangle > 10^{-7}$ (Eqs.\,(\ref{190527a}) and (\ref{190523z})) that extends the size of the chaotic zone (Fig.\,\ref{SSKOI}, top).
As a result, the rotation may remain permanently chaotic or evolve into some totally unexpected spin-orbit resonance (see next section for more details).

\section{Complete evolution}
\llabel{evol}

The spin evolution of a planet is governed by equation (\ref{130104d}).
For a preliminary analysis, this equation can be decoupled in the shape of the planet $Z$ (Eq.\,(\ref{190221a})) and in the orbital forcing function $x$ (Eq.\,(\ref{190221b})). 
The orbital forcing determines the equilibrium states for the spin, and for a given shape we can predict the possible dynamical regimes.
However, when we take into account tidal deformation, the shape itself depends on the orbital forcing (Eq.\,(\ref{190508a})).
As a result, the exact behavior of the spin depends on the tidal model that we adopt.

From the analysis of the orbital forcing, we can anticipate which systems can present a rich spin dynamics, such as the KOI-1599 system.
However, a deep understanding of all possible scenarios can only be achieved by performing numerical simulations of the complete equations of motion, namely the orbital equations of an $N$-body system, which give the solution for $x(t)$, together with equation (\ref{130104d}) for the spin and equation (\ref{eq:Z}) for the deformation (tidal model).
At the end of the previous section, we already did that for estimating the numerical probabilities of capture in a specific spin-orbit resonance (Figs.~\ref{PcapM} and \ref{PcapKOI}).

In this section, we study the complete behavior of the outer planet in the KOI-1599 system (Table~\ref{TkoiGEAI}).
This planet has an estimated mass of 4.3~$M_\oplus$, that is, it can be classified as a super-Earth. 
An upper bound of ten Earth masses is commonly accepted to warrant that the planet is mostly rocky \citep[e.g.,][]{Valencia_etal_2007a}.
Therefore, the internal structure and rheology of this planet is likely similar to the terrestrial planets in the solar system.
We adopt for the residual deformation $Z_0 = 1.0 \times 10^{-7}$. 
For the tidal deformation we use a Maxwell model (Eq.\,(\ref{eq:Z})), with  
$\kf = 0.9$, $\taua = 0$, and $C/\m\R^2 = 1/3$. 

For rocky planets, such as the Earth and Mars, we have a dissipation factor $Q=10$ \citep{Dickey_etal_1994} and $Q=80$ \citep{Lainey_etal_2007}, respectively.
We then compute for the Earth $\taub=1.6$~day, and for Mars $\taub=14.7$~day.
However, in the case of the Earth, the present $Q-$factor is dominated by the oceans, the Earth's solid body $Q$ is estimated to be 280 \citep{Ray_etal_2001}, which increases the relaxation time by more than one order of magnitude ($\taub=46$~day).
Although these values provide a good estimate for the average present dissipation ratios, they appear to be inconsistent with the observed deformation of the planets.
Indeed, in the case of the Earth, the surface post-glacial rebound from the last glaciation about $10^4$~years ago is still going on, suggesting that the Earth's mantle relaxation time is something like $\taub=4400$~yr \citep{Turcotte_Schubert_2002}.
Since $\taub$ can range from a few minutes up to thousands of years,
we adopt $- 6 \le \log_{10} \taub_\mathrm{yr} \le 4$, which is equivalent to $ - 4 \le \log_{10} \taub n \le 6 $.

In our numerical simulations we use the initial conditions from Table~\ref{TkoiGEAI}.
The initial rotation period is set at 19.5~day, which corresponds to $\dot \theta / n \approx 1.05 $.
Since the libration frequency of the outer planet is $\lbf/n \approx 0.0066$ (Table~\ref{TkoiAF}), the initial rotation rate is completely outside the resonant area, which is bounded by $\dot \theta/n \lesssim 1.01$ (Fig.~\ref{SSKOI}).
For any $\tau$ value, the rotation rate of the planet decreases owing to tidal effects until it approaches the region where multiple spin-orbit resonances are present.
The shape of the tidal torque depends on the $\taub$ value (Fig.~\ref{NetTorque}), so at this stage different scenarios can occur.
In addition, capture in resonance is a stochastic process.
Therefore, for each $\taub$ value we have run 180 simulations with $\theta_0$ differing by one degree.
In Table~\ref{tabprobs} we list the final distribution of the spin in the different resonances for each $\taub$ value.

\begin{table*}
\begin{center}
\caption{Number of captures in each spin-orbit resonances (in percentage) for the KOI-1599 outer planet using different $\tau$ values. \label{tabprobs}}
\begin{tabular}{l | r | r | r | r | r | r | r | r | r | r | r}
 $\tau n$ &  $10^{-4}$  &  $10^{-3}$  &  $10^{-2}$  & $10^{-1}$  & $10^0$  &  $10^1$  &  $10^2$ & $10^3$ & $10^4$ & $10^5$ & $10^6$  \\
\hline
$n+\lbf   $ &   5.1 &   9.4 &   3.3 &    $-$ &   $-$ &  $-$ &  $-$ &  $-$  &   $-$ &  $-$ &  $-$  \\
$n+\lbf/2$ & 64.0 & 47.5 & 58.3 &  74.0 &  10.6 &  $-$ &  $-$ &  $-$  & 59.2 &  $-$ &  $-$  \\
$n$           & 30.3 & 39.2 & 36.7 &  26.0 &  89.4 & 100 & 100 & 100  &   $-$ &  $-$ &  $-$  \\
$n-\lbf/2$  &   0.6 &   3.9 &   1.7 &    $-$ &   $-$ &  $-$ &  $-$ &  $-$  & 40.8 &  $-$ &  $-$  \\
$n-\lbf$     &   $-$ &   $-$ &  $-$ &    $-$ &    $-$ &  $-$ &  $-$ &  $-$  & $-$  &  $-$ &  86.7  \\
chaotic      &   $-$ &  $-$ &   $-$ &   $-$ &    $-$ &  $-$ &  $-$ &  $-$  &  $-$  & 100 & 13.3 \\
\hline
\end{tabular}
\end{center}
\end{table*}

\begin{figure}
    \centering
    \includegraphics[width=\columnwidth]{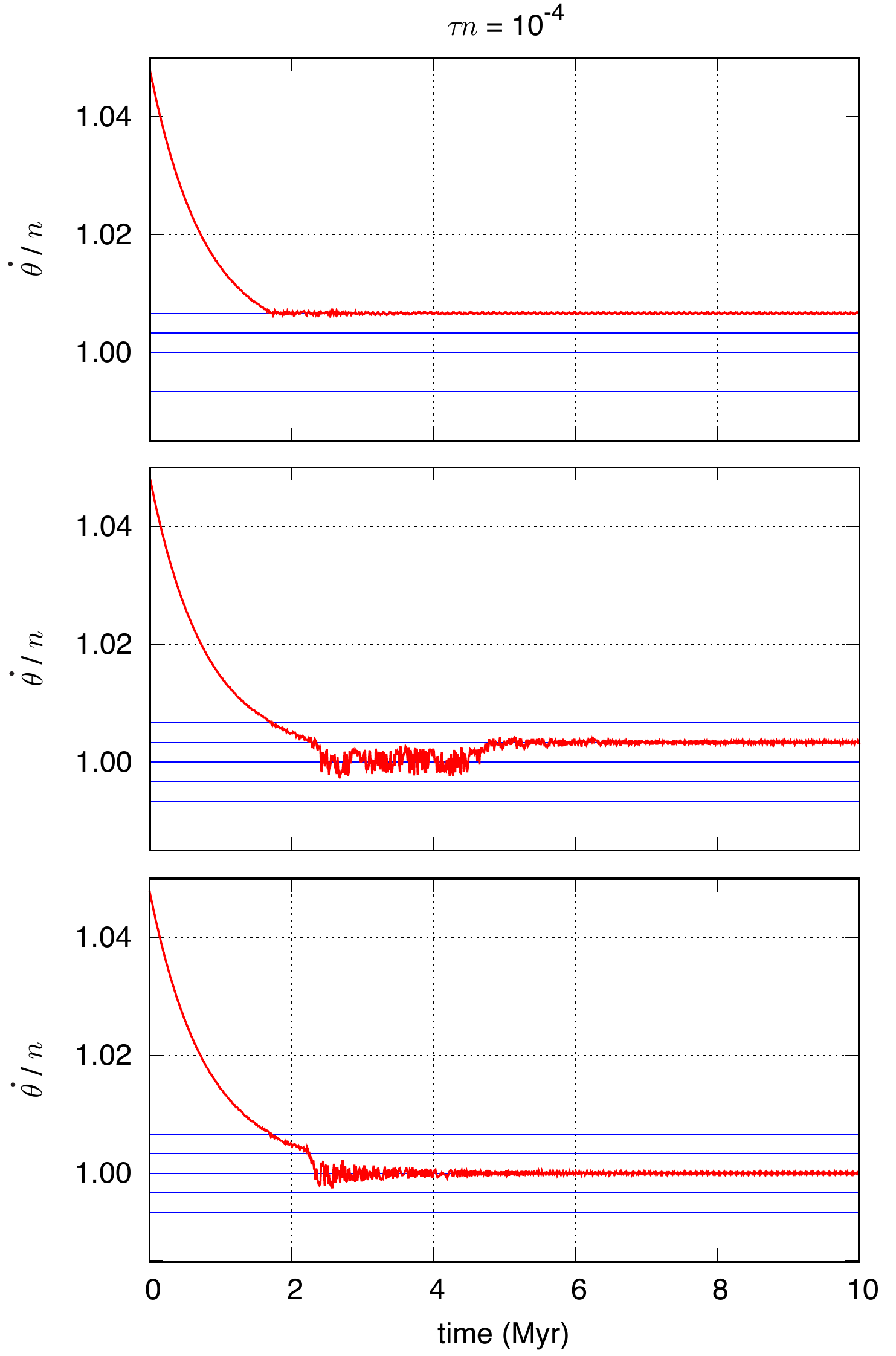}
    \caption{Some examples of the spin evolution of the outer planet in the KOI-1599 system for $\taub n = 10^{-4}$ (low frequency regime). The blue lines give the position of the main spin-orbit resonances $n$, $n\pm\lbf/2$, and $n\pm\lbf$.}
    \label{tau3}
 \end{figure}

For $\taub n \ll 10^{2}$, the spin is in the ``low frequency regime'' ($\taub \lbf \ll 1$), which is usually known as the viscous or linear regime \citep[e.g.,][]{Singer_1968, Mignard_1979}.
Because the relaxation time is small, the deformation of the planet is very close to the equilibrium deformation, $Z \approx Z_0 + Z_e$ (Eq.\,(\ref{eq:Z})).
As a consequence, $\langle Z \rangle \approx Z_0$, that is, the behavior of the spin follows that of a rigid body (section~\ref{rigid}).
We can thus predict the dynamics from the frequency map analysis shown in Figure~\ref{SSKOI} (bottom). 
We see that the rotation can be trapped in multiple spin-orbit resonances. 
Close to the synchronous resonance, there is a small chaotic zone where the spin can wander for some time before it finds a path into a stable resonant island.
In Figure~\ref{tau3} we show three examples of capture in different spin-orbit resonances.
From Table~\ref{tabprobs} we conclude that the most likely resonance is the super-synchronous resonance with $\omg = n + \lbf/2$.
In the linear regime, the value of $\taub$ only impacts the damping timescale, so the statistics do not change much with this parameter.

For $\tau n \sim 10^2$, the spin is in a ``frequency transition regime'' ($\tau \lbf  \sim 1$). 
In this case, the deformation of the planet is still $\langle Z \rangle \approx Z_0$, but the tidal torque has an important modification.
In the low frequency regime it decreases linearly, allowing capture in all spin-orbit resonances. 
In the transition regime, however, the torque drives the rotation away from all spin-orbit equilibria except for $|\dot \theta / n -1| < 0.002$, where there is an inversion (Fig.~\ref{NetTorque}).
As a result, only the synchronous resonance can be reached (Table~\ref{tabprobs}).
The impact of this inversion in the tidal torque is also pretty clear in Figure~\ref{PcapKOI}, the capture probability in the super-synchronous resonance sharply decreases for $\tau n \sim 10^2$.

For $\taub n \gg 10^2$, the spin is in the ``high frequency regime'' ($\taub \lbf \gg 1$).
In this regime, the relaxation time is large and the planet is thus able to retain some tidal deformation $\langle Z_t \rangle \ne 0 $ (section~\ref{forced}).
For the outer planet in the KOI-1599 system we have $\A = 2.7 \times 10^{-6} \gg Z_0$, so for the main spin-orbit resonances the tidal deformation dominates over the permanent deformation.
As a result, the phase space of the spin is modified with respect to the phase space drawn for $Z_0 = 10^{-7}$, it resembles more to that for $Z_0 \sim 10^{-6}$ (Fig.\,\ref{SSKOI}, top).
We therefore expect chaotic motion for the spin in this regime.
However, in our simulations we verify that, although chaotic rotation is indeed a possibility for some $\taub$ values, stable rotation is also possible in some asynchronous spin-orbit resonances (Table~\ref{tabprobs}).
One reason for this is that in this regime the tidal torque facilitates the capture in all these resonances (Fig.~\ref{NetTorque}).
The other reason is because the tidal deformation varies from resonance to resonance (Eq.\,(\ref{190417c})).
For the synchronous resonance $| \langle Z_t \rangle | \sim 10^{-6}$, so even if the rotation is somehow captured there, it becomes chaotic after some time.
For the super-synchronous resonance $|\langle Z_t \rangle | \approx 3.6 \times 10^{-7}$ (Eq.\,(\ref{190523c})), so it is not impossible to remain trapped there, provided that the spin finds a way in.

\begin{figure}
    \centering
    \includegraphics[width=\columnwidth]{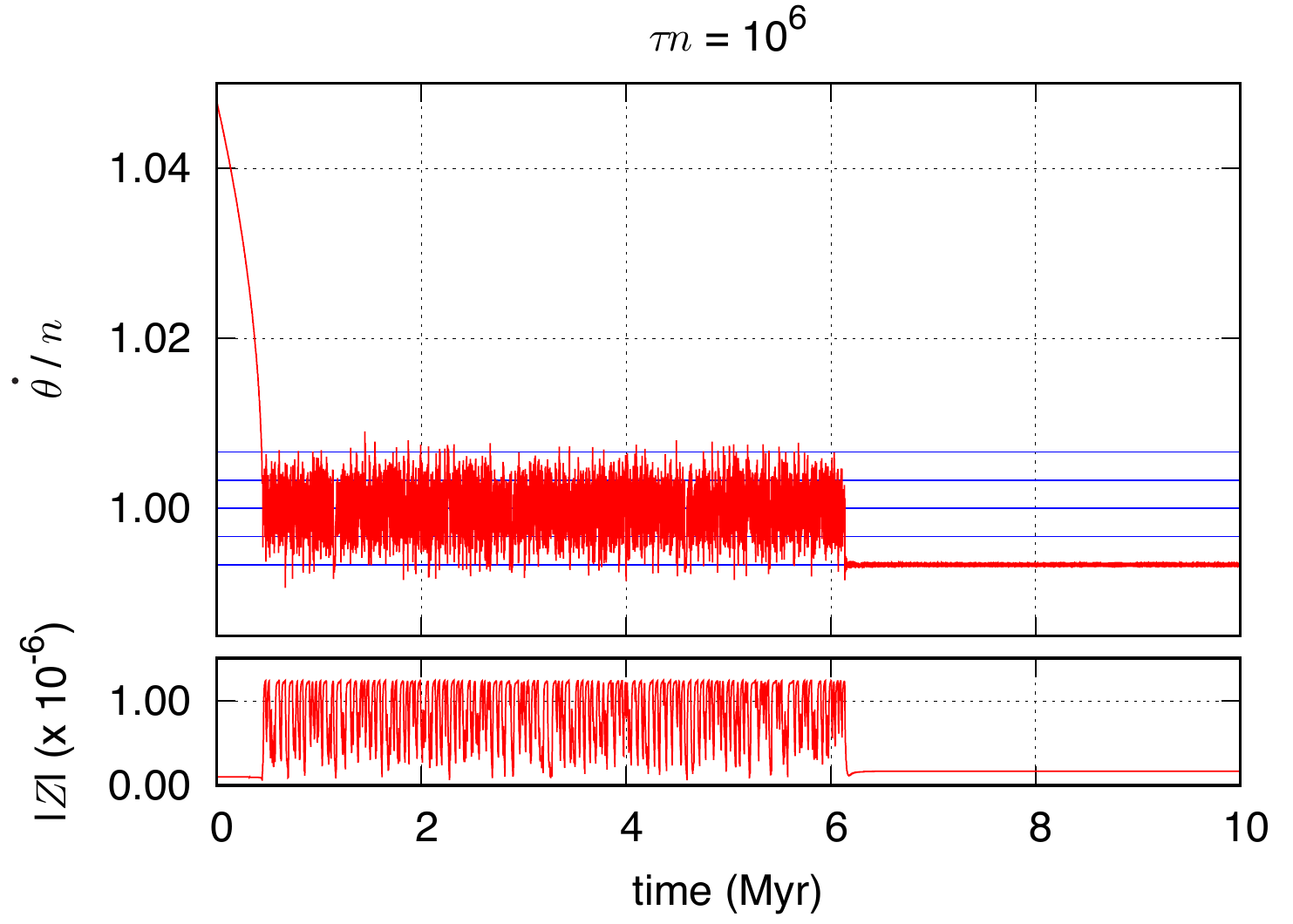}
    \caption{Spin evolution of the outer planet in the KOI-1599 system (top) and the corresponding deformation $|Z|$ (bottom) for $\taub n=10^6$ (high frequency regime). The blue lines give the position of the main spin-orbit resonances $n$, $n\pm\lbf/2$, and $n\pm\lbf$.}
    \label{tau6}
  \end{figure}

In Figure~\ref{tau6} we show an example for $\taub n = 10^6$, which clarifies the behavior of the spin in the high frequency regime.
In this figure we show the evolution of the rotation rate (top) together with the evolution of the shape (bottom).
Before the spin reaches the resonant area, we have $\langle Z \rangle = Z_0 = 10^{-7}$, so in principle the rotation can be trapped in a stable spin-orbit resonance (Fig.~\ref{SSKOI}, bottom).
As soon as the spin enters the resonant area, the deformation increases and can experience values  $|Z| > 10^{-6}$.
Then, the rotation becomes chaotic (Fig.~\ref{SSKOI}, top).
However, just outside the chaotic zone there exists the sub-synchronous resonance with $\omg = n - \lbf$.
Somehow, after some time in the chaotic region, the spin finds a path into this stable region.
Once there, the deformation of the planet decreases again to a value $|Z| \sim 10^{-7}$.
Thus, the sub-synchronous resonance is no longer connected with the chaotic region and the spin can remain there forever (Fig.~\ref{SSKOI}, bottom).
In this example, it is remarkable that although the spin starts with an initial rotation rate faster than the synchronous rotation, it finally ends up with a rotation smaller than that.

\section{Discussion}
\label{concdisc}

In this paper we have revisited the spin-orbit coupling of close-in planets.
These bodies are assumed to be tidally evolved, and thus can be trapped in some equilibrium states.
We derived our results using a very general approach, that is suitable to apply to multi-planet systems in various orbital configurations (e.g., very eccentric orbits or involved in orbital resonances).
We introduced a simple way to probe the spin dynamics from the orbital perturbations (Eq.\,(\ref{150618a})), a new method for computing the forced librations (Eq.\,(\ref{190507a})) and the permanent tidal deformation (Eq.\,(\ref{190523c})), and a general expression for the tidal torque (Eq.\,(\ref{190507c})) and for the capture probabilities in resonance (Eq.\,(\ref{190627a})). 

We considered that the spin of the planet is only subject to tidal torques from the parent star (Eq.(\ref{130104d})).
Tidal torques between planets are also present, but their contribution is very small for the main spin-orbit resonances \citep[see Table~II in][]{Goldreich_Peale_1966}, and accounts for less than a few percent of the total tidal dissipation \citep{Hay_Matsuyama_2019}.
In addition, the positions of the spin-orbit resonances do not change because they depend on the fundamental frequencies of the system; the additional torques only modify the libration width of the resonant islands.
However, if we consider a system with two or more stars, our model should be corrected to include the torques from all bodies on the spin \citep[][]{Correia_etal_2015, Leleu_2017}.

In our model we assumed for simplicity that the spin axis of the planet is normal to the orbit, since this is the final outcome of tidal evolution \citep{Hut_1980}.
For multi-planet systems, the obliquity does not evolve exactly to zero degrees, it is rather captured in an equilibrium configuration known as Cassini state \citep{Colombo_1966, Peale_1969}.
Although high obliquity states exist, they are usually unstable under strong tidal interactions.
Therefore, the only stable equilibrium state has in general very small obliquity (less than $1^\circ$), for which our model applies.
We note, however, that this may not be the case for planetary systems showing large mutual inclinations \citep{Correia_2015}.

We also assumed coplanar orbits.
This restriction was used to simplify the orbital forcing function $x$ (Eq.(\ref{190221d})), since this function has only two dimensions in the planar case.
If the system is not coplanar we can still use our model because the forcing function captures the effect of the inclination.
It is nevertheless convenient to align the inertial plane where we compute the forcing function with the mean orbital plane of the planet whose rotation we want to study.
This alignment is also important to ensure that the obliquity remains close to zero.

In our model we only considered gravitational tidal torques.
For Earth-size or larger planets, we can expect that a dense atmosphere is present.
As a result, thermal atmospheric tides may give rise to additional equilibrium states \citep{Correia_Laskar_2001}.
However, it has been shown that gravitational tides dominate for close-in planets with $a \lesssim 0.3$~au \citep{Correia_etal_2008, Leconte_etal_2015}.
Thus, as long as we stay in this regime, the presence of an atmosphere should not modify the global picture drawn in this paper much.

We have shown that the spin dynamics of Earth-size planets in compact systems can be very rich.
In particular, we have shown that synchronous rotation is not the only possibility; the rotation is allowed to evolve into some unexpected spin-orbit resonances or even present chaotic behavior.
In the high frequency regime, we have shown that the tidal torque has equilibrium points at exactly the same position as the spin-orbit resonances, which enhances the chances of capture in these equilibria.
Future studies can take into account all the neglected contributions in our model, which may increase the diversity of spin states for these planets even more.

\begin{acknowledgements}
We thank G.~Bou\'e, A.~Leleu, and P.~Robutel for discussions. 
AC acknowledges support by 
CFisUC strategic project (UID/FIS/04564/2019),
ENGAGE SKA (POCI-01-0145-FEDER-022217), and
PHOBOS (POCI-01-0145-FEDER-029932),
funded by COMPETE 2020 and FCT, Portugal.
JBD acknowledges support by the Swiss National Science Foundation (SNSF).
This work has, in part, been carried out within the framework of
the National Centre for Competence in Research PlanetS
supported by SNSF.
\end{acknowledgements}


\bibliographystyle{aa}
\bibliography{\bibpath correia}

\begin{thebibliography}{60}
\expandafter\ifx\csname natexlab\endcsname\relax\def\natexlab#1{#1}\fi

\bibitem[{{Adams} \& {Bloch}(2015)}]{Adams_Bloch_2015}
{Adams}, F.~C. \& {Bloch}, A.~M. 2015, \mnras, 446, 3676

\bibitem[{{Andrade}(1910)}]{Andrade_1910}
{Andrade}, E.~N.~d.~C. 1910, Proc. R. Soc. Lond. A, 84, 1

\bibitem[{{Carpenter}(1964)}]{Carpenter_1964}
{Carpenter}, R.~L. 1964, \aj, 69, 2

\bibitem[{{Chirikov}(1979)}]{Chirikov_1979}
{Chirikov}, B.~V. 1979, Physics Reports, 52, 263

\bibitem[{{Colombo}(1965)}]{Colombo_1965}
{Colombo}, G. 1965, \nat, 208, 575

\bibitem[{{Colombo}(1966)}]{Colombo_1966}
{Colombo}, G. 1966, \aj, 71, 891

\bibitem[{{Correia}(2015)}]{Correia_2015}
{Correia}, A.~C.~M. 2015, \aap, 582, A69

\bibitem[{{Correia}(2018)}]{Correia_2018}
{Correia}, A.~C.~M. 2018, \icarus, 305, 250

\bibitem[{{Correia} {et~al.}(2014){Correia}, {Bou{\'e}}, {Laskar}, \&
  {Rodr{\'{\i}}guez}}]{Correia_etal_2014}
{Correia}, A.~C.~M., {Bou{\'e}}, G., {Laskar}, J., \& {Rodr{\'{\i}}guez}, A.
  2014, \aap, 571, A50

\bibitem[{{Correia} \& {Laskar}(2001)}]{Correia_Laskar_2001}
{Correia}, A.~C.~M. \& {Laskar}, J. 2001, \nat, 411, 767

\bibitem[{{Correia} \& {Laskar}(2004)}]{Correia_Laskar_2004}
{Correia}, A.~C.~M. \& {Laskar}, J. 2004, \nat, 429, 848

\bibitem[{{Correia} \& {Laskar}(2010)}]{Correia_Laskar_2010}
{Correia}, A.~C.~M. \& {Laskar}, J. 2010, \icarus, 205, 338

\bibitem[{{Correia} {et~al.}(2015){Correia}, {Leleu}, {Rambaux}, \&
  {Robutel}}]{Correia_etal_2015}
{Correia}, A.~C.~M., {Leleu}, A., {Rambaux}, N., \& {Robutel}, P. 2015, \aap,
  580, L14

\bibitem[{{Correia} \& {Robutel}(2013)}]{Correia_Robutel_2013}
{Correia}, A.~C.~M. \& {Robutel}, P. 2013, \apj, 779, 20

\bibitem[{{Correia} \& {Rodr{\'{\i}}guez}(2013)}]{Correia_Rodriguez_2013}
{Correia}, A.~C.~M. \& {Rodr{\'{\i}}guez}, A. 2013, \apj, 767, 128

\bibitem[{{Correia} {et~al.}(2008){Correia}, {Udry}, {Mayor}, {Eggenberger},
  {Naef}, {Beuzit}, {Perrier}, {Queloz}, {Sivan}, {Pepe}, {Santos}, \&
  {S{\'e}gransan}}]{Correia_etal_2008}
{Correia}, A.~C.~M., {Udry}, S., {Mayor}, M., {et~al.} 2008, \aap, 479, 271

\bibitem[{{Darwin}(1879)}]{Darwin_1879a}
{Darwin}, G.~H. 1879, Philos. Trans. R. Soc. London, 170, 1

\bibitem[{{Darwin}(1880)}]{Darwin_1880}
{Darwin}, G.~H. 1880, Philos. Trans. R. Soc. London, 171, 713

\bibitem[{{Delisle} {et~al.}(2017){Delisle}, {Correia}, {Leleu}, \&
  {Robutel}}]{Delisle_etal_2017}
{Delisle}, J.-B., {Correia}, A.~C.~M., {Leleu}, A., \& {Robutel}, P. 2017,
  \aap, 605, A37

\bibitem[{{Dickey} {et~al.}(1994){Dickey}, {Bender}, {Faller}, {Newhall},
  {Ricklefs}, {Ries}, {Shelus}, {Veillet}, {Whipple}, {Wiant}, {Williams}, \&
  {Yoder}}]{Dickey_etal_1994}
{Dickey}, J.~O., {Bender}, P.~L., {Faller}, J.~E., {et~al.} 1994, Science, 265,
  482

\bibitem[{{Dobrovolskis}(1980)}]{Dobrovolskis_1980}
{Dobrovolskis}, A.~R. 1980, \icarus, 41, 18

\bibitem[{{Dones} \& {Tremaine}(1993)}]{Dones_Tremaine_1993}
{Dones}, L. \& {Tremaine}, S. 1993, \icarus, 103, 67

\bibitem[{{Efroimsky}(2012)}]{Efroimsky_2012}
{Efroimsky}, M. 2012, Celestial Mechanics and Dynamical Astronomy, 112, 283

\bibitem[{{Genova} {et~al.}(2013){Genova}, {Iess}, \&
  {Marabucci}}]{Genova_etal_2013}
{Genova}, A., {Iess}, L., \& {Marabucci}, M. 2013, \planss, 81, 55

\bibitem[{{Gold} \& {Soter}(1969)}]{Gold_Soter_1969}
{Gold}, T. \& {Soter}, S. 1969, \icarus, 11, 356

\bibitem[{{Goldreich} \& {Peale}(1966)}]{Goldreich_Peale_1966}
{Goldreich}, P. \& {Peale}, S. 1966, \aj, 71, 425

\bibitem[{{Goldreich} \& {Peale}(1967)}]{Goldreich_Peale_1967}
{Goldreich}, P. \& {Peale}, S. 1967, \aj, 72, 662

\bibitem[{{Goldreich} \& {Soter}(1966)}]{Goldreich_Soter_1966}
{Goldreich}, P. \& {Soter}, S. 1966, \icarus, 5, 375

\bibitem[{{Goldstein}(1964)}]{Goldstein_1964}
{Goldstein}, R.~M. 1964, \aj, 69, 12

\bibitem[{{Harbison} {et~al.}(2011){Harbison}, {Thomas}, \&
  {Nicholson}}]{Harbison_etal_2011}
{Harbison}, R.~A., {Thomas}, P.~C., \& {Nicholson}, P.~C. 2011, Celestial
  Mechanics and Dynamical Astronomy, 110, 1

\bibitem[{{Hay} \& {Matsuyama}(2019)}]{Hay_Matsuyama_2019}
{Hay}, H. C.~F.~C. \& {Matsuyama}, I. 2019, \apj, 875, 22

\bibitem[{{Henning} {et~al.}(2009){Henning}, {O'Connell}, \&
  {Sasselov}}]{Henning_etal_2009}
{Henning}, W.~G., {O'Connell}, R.~J., \& {Sasselov}, D.~D. 2009, \apj, 707,
  1000

\bibitem[{{Henrard}(1993)}]{Henrard_1993}
{Henrard}, J. 1993, in Dynamics Reported (Springer Verlag, New York), 117--235

\bibitem[{{Hughes}(1981)}]{Hughes_1981}
{Hughes}, S. 1981, Celestial Mechanics, 25, 101

\bibitem[{{Hut}(1980)}]{Hut_1980}
{Hut}, P. 1980, \aap, 92, 167

\bibitem[{{Kokubo} \& {Ida}(2007)}]{Kokubo_Ida_2007}
{Kokubo}, E. \& {Ida}, S. 2007, \apj, 671, 2082

\bibitem[{{Lainey} {et~al.}(2007){Lainey}, {Dehant}, \&
  {P{\"a}tzold}}]{Lainey_etal_2007}
{Lainey}, V., {Dehant}, V., \& {P{\"a}tzold}, M. 2007, \aap, 465, 1075

\bibitem[{{Laskar}(1993)}]{Laskar_1993PD}
{Laskar}, J. 1993, Physica D Nonlinear Phenomena, 67, 257

\bibitem[{{Leconte} {et~al.}(2015){Leconte}, {Wu}, {Menou}, \&
  {Murray}}]{Leconte_etal_2015}
{Leconte}, J., {Wu}, H., {Menou}, K., \& {Murray}, N. 2015, Science, 347, 632

\bibitem[{{Leleu}(2017)}]{Leleu_2017}
{Leleu}, A. 2017, arXiv e-prints, arXiv:1701.05585

\bibitem[{{Leleu} {et~al.}(2016){Leleu}, {Robutel}, \&
  {Correia}}]{Leleu_etal_2016}
{Leleu}, A., {Robutel}, P., \& {Correia}, A.~C.~M. 2016, Celestial Mechanics
  and Dynamical Astronomy, 125, 223

\bibitem[{{MacDonald}(1964)}]{MacDonald_1964}
{MacDonald}, G.~J.~F. 1964, Revs. Geophys., 2, 467

\bibitem[{{Mignard}(1979)}]{Mignard_1979}
{Mignard}, F. 1979, Moon and Planets, 20, 301

\bibitem[{{Morbidelli}(2002)}]{Morbidelli_2002}
{Morbidelli}, A. 2002, {Modern celestial mechanics : aspects of solar system
  dynamics} (London: Taylor \& Francis)

\bibitem[{{Murray} \& {Dermott}(1999)}]{Murray_Dermott_1999}
{Murray}, C.~D. \& {Dermott}, S.~F. 1999, {Solar System Dynamics} (Cambridge
  University Press)

\bibitem[{{Noyelles} {et~al.}(2014){Noyelles}, {Frouard}, {Makarov}, \&
  {Efroimsky}}]{Noyelles_etal_2014}
{Noyelles}, B., {Frouard}, J., {Makarov}, V.~V., \& {Efroimsky}, M. 2014,
  \icarus, 241, 26

\bibitem[{{Panichi} {et~al.}(2019){Panichi}, {Migaszewski}, \&
  {Go{\'z}dziewski}}]{Panichi_etal_2019}
{Panichi}, F., {Migaszewski}, C., \& {Go{\'z}dziewski}, K. 2019, \mnras, 485,
  4601

\bibitem[{{Peale}(1969)}]{Peale_1969}
{Peale}, S.~J. 1969, \aj, 74, 483

\bibitem[{{Peale} {et~al.}(2007){Peale}, {Yseboodt}, \&
  {Margot}}]{Peale_etal_2007}
{Peale}, S.~J., {Yseboodt}, M., \& {Margot}, J.-L. 2007, \icarus, 187, 365

\bibitem[{{Pettengill} \& {Dyce}(1965)}]{Pettengill_Dyce_1965}
{Pettengill}, G.~H. \& {Dyce}, R.~B. 1965, \nat, 206, 1240

\bibitem[{{Ray} {et~al.}(2001){Ray}, {Eanes}, \& {Lemoine}}]{Ray_etal_2001}
{Ray}, R.~D., {Eanes}, R.~J., \& {Lemoine}, F.~G. 2001, Geophysical Journal
  International, 144, 471

\bibitem[{{Showalter} \& {Hamilton}(2015)}]{Showalter_Hamilton_2015}
{Showalter}, M.~R. \& {Hamilton}, D.~P. 2015, \nat, 522, 45

\bibitem[{{Singer}(1968)}]{Singer_1968}
{Singer}, S.~F. 1968, \gjras, 15, 205

\bibitem[{{Smith} {et~al.}(1982){Smith}, {Soderblom}, {Batson}, {Bridges},
  {Inge}, {Masursky}, {Shoemaker}, {Beebe}, {Boyce}, {Briggs}, {Bunker},
  {Collins}, {Hansen}, {Johnson}, {Mitchell}, {Terrile}, {Cook}, {Cuzzi},
  {Pollack}, {Danielson}, {Ingersoll}, {Davies}, {Hunt}, {Morrison}, {Owen},
  {Sagan}, {Veverka}, {Strom}, \& {Suomi}}]{Smith_etal_1982}
{Smith}, B.~A., {Soderblom}, L., {Batson}, R.~M., {et~al.} 1982, Science, 215,
  504

\bibitem[{{Steffen} \& {Hwang}(2015)}]{Steffen_Hwang_2015}
{Steffen}, J.~H. \& {Hwang}, J.~A. 2015, \mnras, 448, 1956

\bibitem[{{Thomas} {et~al.}(2007){Thomas}, {Armstrong}, {Asmar}, {Burns},
  {Denk}, {Giese}, {Helfenstein}, {Iess}, {Johnson}, {McEwen}, {Nicolaisen},
  {Porco}, {Rappaport}, {Richardson}, {Somenzi}, {Tortora}, {Turtle}, \&
  {Veverka}}]{Thomas_etal_2007}
{Thomas}, P.~C., {Armstrong}, J.~W., {Asmar}, S.~W., {et~al.} 2007, \nat, 448,
  50

\bibitem[{{Turcotte} \& {Schubert}(2002)}]{Turcotte_Schubert_2002}
{Turcotte}, D.~L. \& {Schubert}, G. 2002, {Geodynamics}

\bibitem[{{Valencia} {et~al.}(2007){Valencia}, {Sasselov}, \&
  {O'Connell}}]{Valencia_etal_2007a}
{Valencia}, D., {Sasselov}, D.~D., \& {O'Connell}, R.~J. 2007, \apj, 656, 545

\bibitem[{{Wisdom} {et~al.}(1984){Wisdom}, {Peale}, \&
  {Mignard}}]{Wisdom_etal_1984}
{Wisdom}, J., {Peale}, S.~J., \& {Mignard}, F. 1984, \icarus, 58, 137

\bibitem[{{Yoder}(1995)}]{Yoder_1995cnt}
{Yoder}, C.~F. 1995, in Global Earth Physics: A Handbook of Physical Constants
  (American Geophysical Union, Washington D.C), 1--31

\end{thebibliography}

\end{document}